\documentstyle[12pt,aaspp4]{article}  % single spaced
\catcode`\@=11 % This allows us to modify PLAIN macros.
\def\lsim{\mathrel{\mathpalette\@versim<}}
\def\gsim{\mathrel{\mathpalette\@versim>}}
\def\be{\begin{equation}}
\def\ee{\end{equation}}

\def\K{\rm K}
\def\cP{{\cal P}}

\def\mdot{\dot{m}}
\def\Mdot{\dot{M}}

\def\@versim#1#2{\vcenter{\offinterlineskip
        \ialign{$\m@th#1\hfil##\hfil$\crcr#2\crcr\sim\crcr } }}
\catcode`\@=12 % at signs are no longer letters

\eqsecnum
\begin{document}
\date{}  
\title{Electron-Positron Pairs in Hot Accretion Flows and Thin Disk Coronae}
\author{Ann A. Esin\footnote{AXAF Fellow}} 
\affil{Caltech 130-33, Pasadena, CA 91125}
\affil{aidle@tapir.caltech.edu}

\begin{abstract} 
We investigate equilibrium accretion flows dominated by $e^+ e^-$
pairs.  We consider one- and two-temperature accretion disk coronae above a
thin disk, as well as hot optically thin two-temperature accretion
flows without an underlying thin disk; we model the latter in the
framework of advection-dominated accretion flows (ADAFs).  In all
three cases we include equipartition magnetic fields.  We confirm the
previous result that the equilibrium density of pairs in
two-temperature ADAFs is negligible; and show that the inclusion of
magnetic fields and the corresponding synchrotron cooling reduces the
pair density even further.  Similarly, we find that pairs are
unimportant in two-temperature coronae.  Even when the corona has
significantly enhanced heating by direct transfer of viscous
dissipation in the thin disk to the corona, the inefficient Coulomb
coupling between protons and electrons acts as a bottleneck and
prevents the high compactness required for pair-dominated solutions.
Only in the case of a one-temperature corona model do we find
pair-dominated thermal equilibria.  These pair-dominated solutions
occur over a limited range of optical depth and temperature.
\end{abstract} 

\keywords{accretion, accretion disks -- black hole physics -- plasmas --
radiation mechanisms: thermal}

\section{Introduction}
\label{intro}

Observational and theoretical efforts to understand the emission from
high energy X-ray sources (e.g. AGN and black hole X-ray binaries)
have left no doubt about the existence of astrophysical plasmas with
electron temperatures of order $10^9 - 10^{10}\,\K$.  As soon as the
first hot accretion flow model was constructed (Shapiro, Lightman \&
Eardley 1976, hereafter SLE) it was realized that the production and
annihilation of $e^+ e^-$ pairs may play an important role in such
solutions.  Since then, enormous progress has been made in our
understanding of the nature of $e^+ e^-$ equilibria, first in the
context of static, isothermal hot plasma clouds (e.g. Lightman 1982;
Svensson 1982, 1984; Sikora \& Zbyszewska 1986; Kusunose 1987;
Bj\"ornsson \& Svensson 1991a), and later for more realistic accretion
disk models (e.g.  Kusunose \& Takahara 1988, 1989, 1990; Tritz \&
Tsuruta 1989; White \& Lightman 1989, 1990; Bj\"ornsson \& Svensson
1992; Bj\"ornsson et al.  1996; Kusunose \& Mineshige 1996).
 
Until now, most treatments of pair equilibria (with the exception of the
papers by Zdziarski [1986], Kusunose \& Takahara [1989], and Kusunose \&
Zdziarski [1994]) have considered non-magnetic plasmas, in which the
only photon production mechanisms are bremsstrahlung, double Compton
scattering and pair annihilation (several authors include also an
external soft photon source, e.g. Zdziarski et al. [1994]). However,
there exists now a consensus that magnetic fields are universal in
accretion flows and play an important role in their energy balance.  In
the presence of reasonably strong magnetic fields, synchrotron radiation
dominates over any other photon production mechanism at mildly
relativistic temperatures ($T_e \gsim 10^9\,\K$), and the presence of
synchrotron photons can change significantly the results obtained for
non-magnetized plasmas. 

An external source of photons, as in the case of a hot corona above a
thin disk (e.g. Haardt \& Maraschi 1991, hereafter HM), will also
modify the standard $e^+ e^-$ pair equilibria.  Recently, models for
AGN and galactic X-ray binaries based on pair-dominated coronae have
been extensively discussed in the literature (e.g. Liang 1991; Zdziarski et
al. 1994; Skibo et al. 1995; Stern et al.  1995a, 1995b; Poutanen,
Krolik \& Ryde 1997).  However, to our knowledge, most authors have
concentrated on numerical computations of spectra, and have not given
enough attention to more basic questions such as the existence and
properties of pair equilibria for a range of relevant parameters and
assumptions.  Kusunose \& Mineshige (1991) investigated a
two-temperature disk corona, but they explored a relatively narrow
parameter range.  It is also not clear how their results are related
to one-temperature pair-dominated coronae or a popular scenario
(suggested originally by HM) in which part of the disk viscous energy
is dissipated in the corona.  Moreover, since in their model the
corona is heated through its own differential rotation, the dynamics
of the coronal gas must be treated self-consistently, taking into
account the radial energy advection term, especially for the high
accretion rates considered in the paper.

In this paper we systematically investigate the existence and
properties of pair-dominated equilibrium solutions for three different
scenarios: one-temperature static thin disk coronae, heated by the
viscous dissipation of the gravitational energy stored in the disk,
two-temperature coronae, and hot accretion flows, similar to those
introduced by SLE and Ichimaru (1977, see also Narayan \& Yi 1994 and
Abramowicz et al.  1995).  The latter problem was already discussed by
Bj\"ornsson et al.  (1996) and Kusunose \& Mineshige (1996).  Our
treatment is different only in that we include synchrotron cooling in
our calculation, which was left out by the previous authors.  The
detailed discussion of our initial assumptions and basic equations
used for each of the three scenarios are contained in \S\ref{basiceq}.
We find that pair-dominated solutions are easy to obtain only in the
case of one-temperature coronae, while both two-temperature coronae
and hot accretion flows always have negligible pair densities (see
\S\ref{res}).  These results are discussed in \S\ref{disc}.  The main
conclusions are summarized in \S\ref{sum}.

\section{Basic Equations} 
\label{basiceq} 

Following Bj$\ddot{\rm o}$rnsson \& Svensson (1991, 1992) we divide the
problem of finding equilibrium accretion flow solutions in the presence
of $e^+ e^-$ pairs into two main parts, one dealing with radiative
transfer and pair balance in a static plasma cloud, and the other
describing the details of the dynamics and energetics in an accretion
flow.  Bj$\ddot{\rm o}$rnsson \& Svensson (1991) have shown that the
physical properties of a non-magnetized plasma cloud are completely
determined by specifying only two parameters.  These are the proton
optical depth, $\tau_p = n_p \sigma_T H$, and the dimensionless
luminosity, or compactness parameter, $\ell = q^- H^2 \sigma_T/m_e c^3$,
where $H$ is the size of the cloud, $n_p$ is the proton (i.e. ionized
electron) number density, and $q^-$ is the radiative emissivity per unit
volume.  In this (non-magnetic) case, the mapping between the two parts of the
calculation is particularly simple.  The accretion flow is treated as a
sequence of plasma clouds, whose size is equal to the scale height of
the flow.  At each point in the flow one needs only to specify
$\ell$ and $\tau_p$ in terms of the usual accretion parameters: the mass
of the accreting object $m = M/M_{\odot}$, the radius $r=R/R_S$, where
$R_S= 2G M/c^2$, and the accretion rate $\mdot=\Mdot/\Mdot_{Edd}$,
normalized to the Eddington value $\Mdot_{Edd} = 4 \pi G M m_p/(0.1
\sigma_T c)$. 

In this paper, we attempt to treat a more general problem in which the
accretion flow contains strong magnetic fields, which lead to
non-negligible cooling via synchrotron radiation.  We also allow for
external heating and cooling sources which are relevant in a corona
heated through dissipation of the thin disk gravitational energy and
cooled via Comptonization of thin disk radiation.  When these
effects are properly taken into account, in addition to the parameters
$\tau_p$ and $\ell$, one needs to specify also the particle density
$n_p$ or, equivalently, the cloud size $H$, to completely determine
the properties of a plasma cloud.  With this modification, the mapping
between the cloud and flow properties is no longer as straightforward
as in Bj$\ddot{\rm o}$rnsson \& Svensson (1991, 1992), and depends on
the details of the initial assumptions.  Nevertheless, the mapping
method is still a convenient technique for solving the problem of
electron-positron pairs in accretion flows.

\subsection{Pair and Photon Balance in a Hot Plasma Cloud}
\label{ppbal}

Pair equilibrium requires that
\begin{equation} 
\label{pairb} 
(\dot n_+)_{\rm ann} = (\dot n_+)_{e e} + (\dot n_+)_{e p} +
(\dot n_+)_{\gamma \gamma}+ (\dot n_+)_{\gamma e} +(\dot n_+)_{\gamma_p}, 
\end{equation} 
where $n_+$ is the positron number density, and the rate of pair
annihilation on the left hand side is balanced by the sum of pair
production rates via $e^{\pm}$--$e^{\pm}$, $e^{\pm}$--$p$,
$\gamma$--$\gamma$, $\gamma$--$e^{\pm}$, and $\gamma$--$p$ collisions,
respectively, on the right hand side.  We neglect pair escape from the
cloud.  The analytical expressions for the pair annihilation and
electron-electron pair creation rates in a thermal plasma are given by
Svensson (1982) and White \& Lightman (1989).  According to Svensson
(1982), the $e^{\pm}$--$p$ rate can be ignored, since it is generally
negligible compared with the $e^{\pm}$--$e^{\pm}$ rate.  The last three
terms in equation (\ref{pairb}) depend on the photon spectral density
and, therefore, involve the details of the radiative transfer which we
discuss below.

In a magnetized plasma, thermal synchrotron and bremsstrahlung are the
most important photon production mechanisms (in this paper we ignore
double Compton scattering). In addition, there might be an external
source of soft photons, e.g.  blackbody emission from a thin disk
irradiating a hot corona.  The expressions for thermal
bremsstrahlung emissivity are taken from Svensson (1982).  The self-absorbed
synchrotron emission is modeled using the formalism of Mahadevan et al.
(1995) and Narayan \& Yi (1995).  The magnetic field strength, $B$, is
calculated under the assumption that the gas pressure is in
equipartition with the magnetic field pressure in the cloud.  For a
tangled isotropic magnetic field, this implies that 
\be 
\label{bfield}
\frac{B^2}{24 \pi} = n_p m_e c^2 (\theta_p + (1+2 z) \theta_e), \ee
where $\theta_p = k T_p/m_e c^2$ and $\theta_e = k T_e/m_e c^2$ are
dimensionless proton and electron temperatures (both in electron mass
energy units), and $z=n_+/n_p$ is the pair fraction.  The second term
on the right side of equation (\ref{bfield}) takes into account the
pressure from pairs, as well as ionized electrons.

In general, most of the primary photons are too soft to produce pairs.
However, when inverse Compton scattering is important, many of the
soft photons are boosted up to the Wien peak.  In that case collisions
between photons in the Wien peak as well as between Wien and
bremsstrahlung photons can dominate pair-production.  We follow the
approach of Svensson (1984, hereafter S84) and consider the radiation
field as a sum of a 'flat' bremsstrahlung continuum and a Wien
distribution.  In this special case, the photon-particle and
photon-photon pair production rate equations can be integrated
analytically, and the resulting expressions depend only on the number
density of the Wien photons, $n_{\gamma}$.

In order to compute $n_{\gamma}$, we solve the photon balance
equation 
\be 
\label{photonb} 
f_B \dot n_{\gamma}^B + f_S \dot n_{\gamma}^S + f_D \dot n_{\gamma}^D 
= n_{\gamma}/t_{\rm esc}, 
\ee 
where $\dot n_{\gamma}^B$, $\dot n_{\gamma}^S$, and $\dot n_{\gamma}^D$
are the rates of bremsstrahlung, synchrotron, and external soft photon
production respectively, and $f_B$, $f_S$, and $f_D$ represent the
fractions of emitted photons that are scattered into the Wien peak
before escaping from the cloud.  The details of how we compute these
quantities are discussed in Appendix \ref{apa}.  Finally, we approximate
the photon escape time as $t_{\rm esc} = H/c (1+ \tau_T g_W (\theta_e))$
(S84), where $\tau_T = (1+2 z)\tau_p$ is the total scattering optical
depth of the cloud, and the factor $g_W (\theta_e)$ incorporates the
Klein-Nishina effect, averaged over the Wien photon distribution. 

For fixed $\theta_e$, $\tau_p$ and $n_p$, equations (\ref{pairb}),
(\ref{bfield}), and (\ref{photonb}) can be solved to obtain the
equilibrium pair fraction $z=n_+/n_p$ and the cloud compactness,
$\ell$, which in general includes contributions from Comptonized
bremsstrahlung and synchrotron emission, as well as Comptonization of
the external soft radiation (see Appendix \ref{apa}).  As was pointed
out by S84, in non-magnetized plasmas the results are practically
independent of $n_p$.  The only dependence comes from the
determination of the bremsstrahlung self-absorption energy, $x_B$,
defined so that at energies below $x_B$, the local bremsstrahlung
spectrum is a blackbody.  The value of $x_B$ is computed using the
expression derived by S84.  Since the bremsstrahlung energy spectrum
is nearly flat, the total emission, as well as the pair production
rate, are dominated by the high-energy end of the spectrum, and have
at best a logarithmic dependence on $x_B$, and therefore, $n_p$.  In
the presence of magnetic fields, however, the value of the proton
number density becomes more important, because it determines the
strength of the magnetic field (see equation (\ref{bfield})) as well
as the value of the synchrotron self-absorption energy, $x_S$,
computed as described in Esin et al. (1996).  As opposed to the case
of bremsstrahlung radiation, synchrotron emission is generally
dominated by the self-absorbed part, so that the dependence on $x_S$
is no longer negligible.

\begin{figure}
\includegraphics{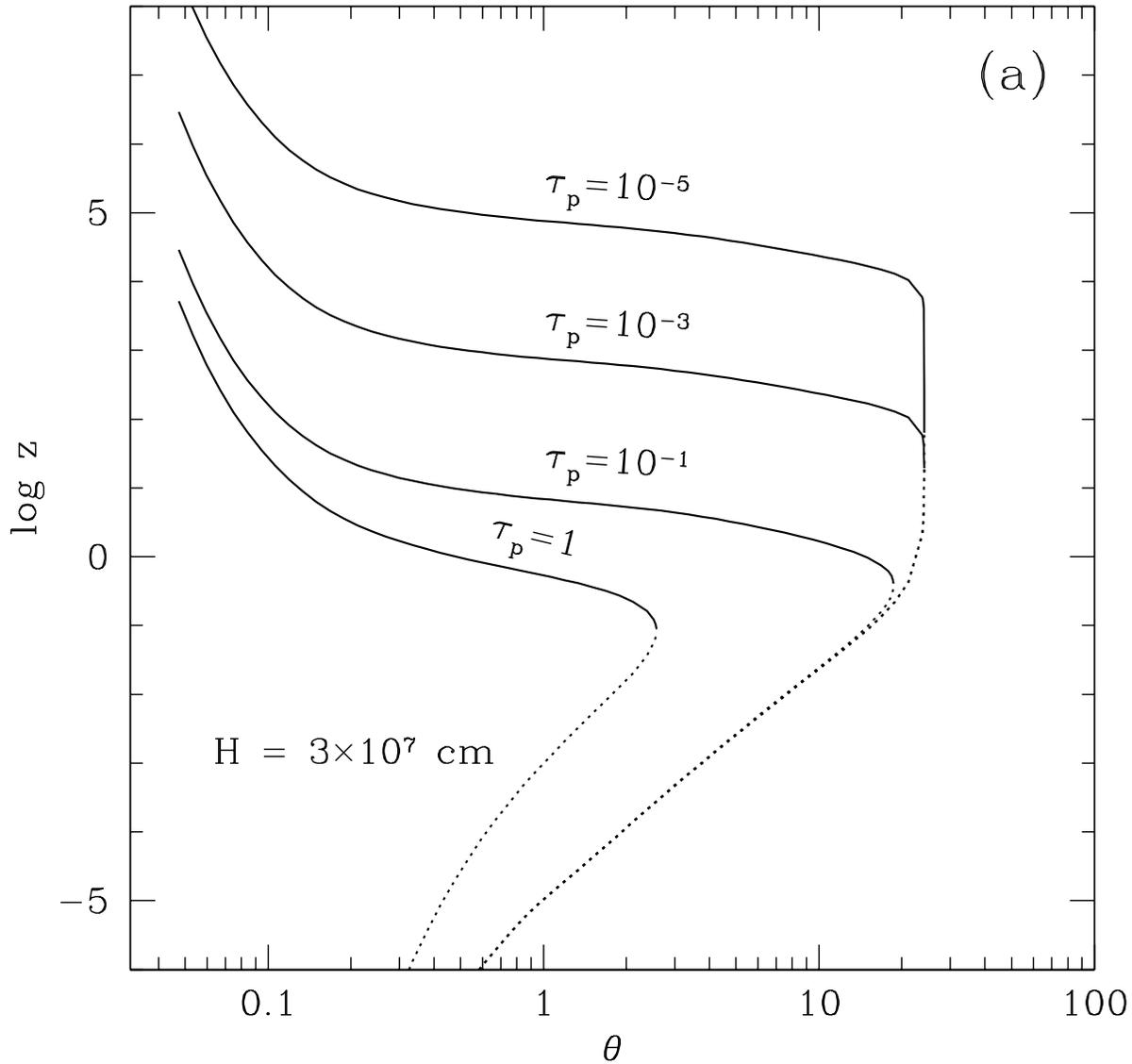} 
\vskip 6.5in 
\caption{\label{zfig} (a) The equilibrium pair fraction, $z = n_+/n_p$, in a
one-temperature plasma cloud cooling via Comptonized bremsstrahlung,
shown as a function of gas temperature.  All four curves were computed
for a cloud of size $H$, but with different values of the proton
optical depth, as indicated on the figure.  Solid lines show the
high-$z$ solution branch and dotted lines show the low-$z$
branch.}
\end{figure}

To compare our calculations with previous work, in Fig.~\ref{zfig} we
plot the equilibrium pair fraction (panel a) and corresponding
compactness (panel b) as functions of the gas temperature $\theta =
\theta_e = \theta_p$ for a single-temperature plasma cloud of a fixed
size.  There is no magnetic field, so that bremsstrahlung is the only
source of soft photons.  Different curves correspond to different
values of $\tau_p$. The results are very similar to those of Svensson
(1982, 1984).  For each value of $\tau_p$, there are two equilibrium
solutions for temperatures below some critical value, $\theta^c
(\tau_p)$; one branch of the solutions (indicated by the solid lines)
is pair-dominated, the other (short-dashed lines) is practically pair-free.
Above $\theta^c$, the pair production rate is greater than the pair
annihilation for any value of $z$ and no equilibrium solution is
possible.  The critical gas temperature is itself a function of the
proton optical depth, increasing with decreasing $\tau_p$, until it
reaches a maximum at $\theta^c \sim 24$.  The pair-dominated solution
branch exists only for $\theta_{BB} < \theta < \theta^c(\tau_p)$,
where $\theta_{BB}$ is the critical temperature at which the pair
fraction reaches its thermal equilibrium value, and the compactness
reaches the blackbody limit (the dashed line in Fig.~\ref{lfig}(b)).

At low $\tau_p$, the equilibrium values of $z$ on the pair-free
solution branch converge to a single curve (Fig.~\ref{zfig}(a)), since
in this regime, pair production is dominated by particle-particle
collisions which have the same scaling with density as the pair
annihilation rate (they both vary as $\tau_p^2$).  The compactness on
the low-$z$ branch is dominated by bremsstrahlung and as a result,
varies as $\tau_p^2$ (Fig.~\ref{lfig}(b)). The high-$z$ branch, on the
other hand, is dominated by photon-photon processes; as a result the
pair density is proportional to $1/\tau_p$, while the
compactness is independent of the proton optical depth, converging to
the Wien equilibrium values for $\theta_e \lsim 0.4$.  The only
significant difference between our results and those of S84 is that
we overestimate the pair fraction in the temperature range $\theta_e
\lsim 0.1$, where double Compton scattering, which we neglect here,
becomes an important source of photons.

Fig.~\ref{zsfig} shows how the pair equilibria in a plasma cloud
change in the presence of magnetic fields.  The magnetic field
strength depends on both the electron and proton temperatures (see
equation (\ref{bfield}), and for simplicity the results shown in
Fig.~\ref{zsfig} were calculated under the assumption that
$\theta_p=\theta_e$.  A comparison of Figs.~\ref{zfig} and \ref{zsfig}
shows that synchrotron cooling significantly changes the shapes of
both $z$ and $\ell$ pair equilibria curves.  On the low-$z$ branch,
the cloud compactness increases strongly for $\theta_e \gsim 0.1$ as
compared to the pure bremsstrahlung case.  Synchrotron cooling causes
the critical temperature to decrease considerably in denser clouds,
with $\tau_p \gsim 10^{-3}$.  By contrast, on the high-$z$ branch the
compactness is practically unchanged, since it follows the Wien
equilibrium curve.  But because of the extra cooling per lepton at
temperatures where synchrotron emission is important ($\theta_e \gsim
0.1$), the equilibrium pair fraction in a magnetic plasma is up to
$10^3$ times smaller than in non-magnetic plasmas.

This figure also illustrates the point made previously, that the
properties of pair equilibria in magnetized plasmas are {\it not}
independent of the cloud size.  For a fixed proton optical depth, a
larger cloud has smaller $n_p$.  Therefore, the value of the
equipartition magnetic field is smaller as well, reducing the
importance of synchrotron cooling.  A comparison between the
short-dashed curve computed for a cloud of size
$H=3\times10^{14}\,{\rm cm}$ and a corresponding curve with the same
value of $\tau_p$ but $H=3\times10^7\,{\rm cm}$ shows this
effect.  It is clear that the dependence on $H$ is not
negligible, though it is not very strong.  The low-$z$ compactness
branch is affected the most, and an increase in $H$ by seven orders of
magnitude, causes the value of $\ell$ to decrease by at most a factor
of $\sim 10^{1.5}$.

\subsection{Energy Balance}
\label{enbal}
To relate the pair equilibrium solutions calculated for static plasma
clouds to a realistic accretion scenario, we must express $n_p$,
$\tau_p$, and $\ell$ in terms of the accretion parameters $\mdot$ and
$r$, and impose an appropriate energy balance condition.  Below we
describe these mapping relations for two different accretion flow
geometries: a corona above a thin disk (\S\S\ref{onetcor},
\ref{twotcor}) and a hot, optically thin accretion flow
(\S\ref{adaf}).

\subsubsection{One-Temperature Disk Coronae}
\label{onetcor}

First we consider the simplest scenario, namely a static, hot,
one-temperature corona above a cool, geometrically thin, optically
thick disk.  The study of such coronae was pioneered by HM.  The
corona is heated through viscous dissipation of the gravitational
energy stored in the thin disk (perhaps via magnetic reconnection,
e.g. Field \& Rogers 1993), and in turn heats the disk by irradiation.
Using the formalism of HM, it is easy to show (see Appendix \ref{apb})
that the total emission per unit area from the thin disk, $Q_D$, is
equal to
\begin{equation}
\label{disken}
Q_D = Q_G \frac{1 - \delta + \delta \eta (1-a)}{1-\eta (1-a) 
(1-\exp{(-\tau_T)})}.
\end{equation}
Here $Q_G = 3 G M \Mdot_D/(8 \pi R^3)$ is the total viscous
dissipation per unit area of the thin disk (e.g. Shakura \& Sunyaev
1973; Frank, King, \& Raine 1992) with mass accretion rate $\Mdot_D$,
$\delta$ is the fraction of viscous energy dissipated directly in the
corona, $\tau_T$ is the total scattering optical depth in the corona,
$\eta$ is the fraction of radiation emitted or scattered in the corona
that is incident on the thin disk, and $a$ is the fraction of incident
radiation that is reflected by the disk.  In general, $\eta$ and $a$
depend on the temperature and optical depth of the corona, the
ionization state of the thin disk, and the spectrum of irradiating
flux.  However, since their exact values are not important in light of
other approximation made in this work, we adopt the values $\eta =
0.5$ and $a = 0.2$ suggested by HM.  This prescription constrains the
disk emission to be always in the range $0.4 Q_G \le Q_D \le 1.7 Q_G$,
regardless of the exact values of $\delta$ and $\tau_T$.  (Note that
$Q_D$ is the disk emission as seen by the corona, not by a remote
observer.) We assume that the thin disk radiates as a black body with
a color temperature
\begin{equation}
\label{td}
T_D = \left(\frac{Q_D}{\sigma_B}\right)^{1/4}.
\end{equation}

To estimate the characteristic size of the corona, we assume that it is in
hydrostatic equilibrium, so that
\begin{equation}
\frac{P}{H} = \rho \frac{G M}{R^2}\ \frac{H}{R} = (m_p n_p + m_e (1+2
z) n_p) \frac{G M}{R^2}\ \frac{H}{R},
\end{equation}  
where $H$ is the scale height of the corona, $\rho$ is the mass
density, and $P$ includes contributions from the magnetic and gas
pressures.  It is interesting to note that though in pair-free plasma
the contribution of electrons to the mass density in the corona is
entirely negligible, if the pair fraction is large enough, namely
$z\gsim m_p/m_e$, pairs must be properly taken into account.  This
allows us to solve for the scale height:
\begin{equation}
\label{h}
h = \frac{H}{R} = \sqrt{\frac{2 r}{\beta}\ \frac{m_e}{m_p}\
\left[\frac{\theta_p + (1+2 z) \theta_e}{1 +(1+2 z)
\frac{m_e}{m_p}}\right]},
\end{equation}
where $(1-\beta)$ is the ratio of the magnetic field pressure to the
total pressure in the corona. In plasmas with no magnetic fields
$\beta=1$.  When magnetic fields are taken into account we always
assume equipartition between the gas and magnetic field pressure,
corresponding to $\beta = 0.5$. 

Knowing $H$, allows us to write the energy balance equation for the
corona, by equating compactness $\ell$ and viscous dissipation (converted
into a dimensionless compactness-like quantity):
\begin{equation}
\label{otcorep}
\ell = \frac{\delta Q_G}{H}\ \frac{H^2 \sigma_T}{m_e c^3} = 
\frac{\delta Q_G H \sigma_T}{m_e c^3}.
\end{equation}
To compute the compactness $\ell$ of the corona, we include
contributions from Comptonized bremsstrahlung and synchrotron
radiation, as well as inverse Compton scattering of the thin disk
photons (see \S\ref{ppbal} and Appendix \ref{apa} for details).

Once the parameters describing the accretion disk ($m$, $r$,
$\mdot_D$) and the corona ($\delta$, $\tau_p$) are fixed, we can solve
equations (\ref{pairb}) through (\ref{otcorep}) to obtain the
equilibrium pair fraction and temperature of the corona.  We begin by
solving pair and photon balance equations for a range of gas
temperatures $\theta = \theta_e = \theta_p$.  As in the case of a
static plasma cloud described in \S\ref{ppbal} above, we obtain two
equilibrium curves for $z (\theta)$ and $\ell (\theta)$, slightly
modified due to the dependence of the scale height $H$ on $\theta$ and
$z$, and the increase in cooling through Comptonization of the thin
disk photons.  Then we calculate the dimensionless viscous dissipation
rate (the right hand side of equation (\ref{otcorep})) as a function
of temperature, again obtaining two solutions for each value of
$\theta$.  Finally, we solve for $\theta$ such that equation
(\ref{otcorep}) is satisfied, i.e. the cooling of the gas in the
corona is balanced by viscous dissipation.  The results are described
in \S\ref{rescor1}.

\subsubsection{Two-Temperature Coronae} 
\label{twotcor} 

Depending on the microphysical details of how viscosity dissipates
gravitational energy, it is feasible that only the protons in the
corona are heated directly (e.g. SLE, Rees et al. 1982).  Since
protons generally cannot radiate, viscous energy must be transfered to
the electrons via Coulomb collisions.  The Coulomb energy transfer is
proportional to the difference between the electron and proton
temperatures, so for the gas to be in thermal equilibrium we must have
protons hotter than electrons, i.e. $\theta_p > \theta_e$.  Such
two-temperature plasmas have been often invoked in the literature in
the context of hot accretion flows (SLE; Rees et al. 1982; Narayan \& Yi
1995; Nakamura, Kusunose, Matsumoto \& Kato 1997; Di Matteo, Blackman
\& Fabian 1997, to give just a few examples) and it is important to
investigate how the results for single-temperature coronae are
modified when proton-electron energy transfer is explicitly taken into
account.

The only difference between one- and two-temperature coronae is that
the latter must satisfy two energy balance equations instead of one.
Energy balance for the protons requires that viscous heating is equal
to the Coulomb energy transfer:
\begin{equation}
\label{ttcorp}
q_{ie} H = \delta Q_G.
\end{equation}
For the electrons to be in thermal equilibrium, we must have
\begin{equation}
\label{ttcore}
\ell = \frac{q_{ie} H^2 \sigma_T}{m_e c^3}.
\end{equation}
The quantity $q_{ie}$ is the rate of Coulomb energy transfer per unit
volume.  In our calculations we use the expression for $q_{ie}$ derived
by Stepney \& Guilbert (1983)
\begin{eqnarray}
\label{qie}
q_{ie} &=& \frac{3}{2} \frac{m_e}{m_p} (1+2 z) n_p^2 \sigma_T c m_e
c^2 \ln{\Lambda} \frac{(\theta_p-\theta_e)}{K_2 (1/\theta_e) K_2
(1/\theta_p^{\prime})} \\ \nonumber
& & \times \left[\frac{2(\theta_e + \theta_p^{\prime})^2 +1}{(\theta_e
+ \theta_P^{\prime})} K_1\left(\frac{\theta_e+\theta_p^{\prime}}{\theta_e
\theta_p^{\prime}}\right) + 2 K_0\left(\frac{\theta_e+\theta_p^{\prime}}
{\theta_e \theta_p^{\prime}}\right)\right],
\end{eqnarray}
where $K_n$ is a modified Bessel function of order $n$,
$\theta_p^{\prime} = \theta_p m_e/m_p$, and the Coulomb logarithm is
taken to be $\ln{\Lambda} = 20$.

To obtain equilibrium solutions for a two-temperature corona, we begin
by solving equations (\ref{pairb})--(\ref{h}), (\ref{ttcorp}), and
(\ref{qie}) for a range of $\theta_e$, to obtain $z(\theta_e)$, $\ell
(\theta_e)$ and $\theta_p (\theta_e)$.  Imposing electron energy
balance then allows us to determine $\theta_e$, such that equation
(\ref{ttcore}) is satisfied.  This procedure is very similar to that
described in \S\ref{onetcor}, except that we have an extra energy
balance equation, which allows us to solve for $\theta_e$ and
$\theta_p$ separately.

\subsubsection{Hot Accretion Flows}
\label{adaf}
In addition to static coronae above a thin disk, we also investigate
pair production in pure optically thin, hot accretion flows.  Such
flows were first proposed by SLE and Ichimaru (1977) and later
extensively studied by Narayan \& Yi (1994, 1995), Abramowicz et al.
(1995), Chen (1995), Chen et al. (1995) and others.  Recently,
Kusunose \& Mineshige (1996) and Bj\"ornsson et al. (1996) addressed
the problem of pairs in the context of these flows.  In this paper we
extend their calculations to include the effects of strong thermal
synchrotron cooling.

As in two-temperature coronae, the protons in hot accretion flows are
heated by viscous dissipation of gravitational energy, which is
parameterized in the usual way through the viscosity parameter
$\alpha$.  However, we allow for the possibility that only a fraction
$(1-f)$ of the dissipated energy is transfered to the electrons via
Coulomb collisions.  The rest of the energy is stored in the gas as
entropy and advected inward with the accretion flow.  When the
accreting object is a black hole, the stored energy is then lost
inside the horizon.  With these assumptions, the energy balance
equation for the protons takes the form
\begin{equation}
\label{adafp}
q_{ie} H = (1-f) Q_G,
\end{equation}
where $Q_G$ is the viscous dissipation rate per unit area of the flow,
while the energy equation for the electrons is the same as in
\S\ref{twotcor} above:
\begin{equation}
\label{adafe}
\ell = \frac{q_{ie} H^2 \sigma_T}{m_e c^3}.
\end{equation}

To compute the rate of viscous dissipation, as well as all other
relevant quantities in the accreting gas (e.g. $n_p$, $H$,
equipartition magnetic field strength $B$, and total pressure $P$), we
use the self-similar solutions derived by Narayan \& Yi (1994, 1995).
These solutions allow us to express various gas properties as
functions of the usual input parameters, $m$, $r$, $\mdot$, and
$\alpha$ as well as the new advection parameter $f$.

As opposed to the two-zone models discussed above, in hot accretion
flows the mass accretion rate $\mdot$, which determines the rate of
viscous dissipation, and the proton optical depth $\tau_p$, which
specifies the cooling rate of the gas, are not independent parameters.
Consequently, to obtain equilibrium accretion flow solutions it is
sufficient to specify $m$, $\mdot$, $r$ and $\alpha$.

We solve for equilibrium solutions in the following manner.  For a
fixed $f$, the scale height of the flow and the characteristic sound
speed at a given radius $r$ are completely determined by the
self-similar solutions.  This allows us to solve for the electron
temperature $\theta_e$ for which the pair balance and electron energy
balance equations, (\ref{pairb})--(\ref{photonb}) and (\ref{adafe}),
are satisfied.  We then solve for the value of $f$ at which the
protons are in thermal equilibrium, determined by equation (\ref{adafp}).

\section{Pair-Dominated Equilibrium Solutions}
\label{res}
The aim of this work is to explore whether equilibrium
pair-dominated accretion flow solutions exist for different flow geometries
and initial assumptions.  We present here the results of
our calculations for the three models described in \S\ref{enbal}.
  
\subsection{One-Temperature Coronae}
\label{rescor1}

Fig.~\ref{corfig} illustrates how equilibrium solutions for a
one-temperature corona above a thin disk are determined.  On it, we
plot the compactness of the corona in pair equilibrium,
$\ell(\theta)$, and the corresponding dimensionless viscous
dissipation rate as functions of the gas temperature $\theta$.  As in
the case of a static plasma cloud, both cooling and heating rates have
two solutions, which we refer to as the high-$z$ and low-$z$ solutions
respectively, for each value of $\theta$.  To emphasize the
distinction between the high-$z$ and low-$z$ solutions, the former are
plotted in Fig.~\ref{corfig} as solid and dashed curves, while the
latter are shown by dotted and dot-dashed curves.  Energy balance
requires that cooling must be balanced by viscous heating; therefore,
the point where the two curves intersect, marked by a circle in the
figure, is an equilibrium solution for the corona.  Because the plot
shows two different solution branches simultaneously, it is important
to remember that valid solutions correspond only to the intersections
of solid lines with dashed lines or dotted lines with dot-dashed
lines.  The other intersections do not give a solution, since the
crossing lines correspond to different values of $z$.

On panel (a) we show our calculations for proton optical depth
$\tau_p=0.1$, while panel (b) corresponds to $\tau_p=10^{-5}$.  The
values of other relevant parameters are given in the figure caption.
Note that the relative slopes of the cooling and heating curves are
such that for any choice of $\mdot$ and $\tau_p$ there is only one
intersection point.  We do not find multiple solutions for any choice
of parameters.

We are interested in pair-dominated solutions, i.e. we wish to find
regions of the parameter space in which the high-$z$ branches of
cooling and viscous heating have an intersection point.  One major
trend is clear already from comparing the two panels in Fig.~\ref{corfig}; 
we see 
that for a fixed mass accretion rate in a thin disk, a corona with a
relatively high proton optical depth is pair-free
(Fig.~\ref{corfig}(a)), while a lower value of $\tau_p$ yields a
pair-dominated solution (Fig.~\ref{corfig}(b)).  This trend is easy to
understand.  The rate of viscous heating is determined by the mass
accretion rate in the disk and is relatively independent of $\tau_p$
in the corona (if we ignore the slight dependence through the vertical
scale height), while the cooling rate on the low-$z$ branch, dominated
by Comptonization of the thin disk radiation, decreases linearly with
$\tau_p$.  Thus, for fixed $\mdot_D$, $r$, and $\delta$, we can find a
critical proton optical depth, $\tau_p^c$, below which the corona is
pair-dominated, i.e has $z \ge 1$.  These maximum values of $\tau_p$
for pair-dominated coronae are plotted in Fig.~\ref{taufig}(a) as a
function of $\mdot_D$.  The solid line shows our results for a plasma
with equipartition magnetic fields.  For comparison, we plot
$\tau_p^c$ for non-magnetized coronae as a dashed line.  The two lines
are practically identical at high $\mdot$, where energy balance in the
corona is dominated by Comptonization of disk photons, but begin to
diverge at lower $\mdot$ where internal cooling in the corona becomes
more important.

The other three panels in Fig.~\ref{taufig} illustrate the dependence of
$\tau_p^c$ on different input parameters.  Fig.~\ref{taufig}(b) shows
how $\tau_p^c$ varies as a function of $\delta$, the fraction of the
gravitational energy of the gas in the thin disk that is dissipated in
the corona.  Lower $\delta$ means that for the same value of $\mdot_D$
the corona is heated less.  To compensate for this, we are driven to
lower values of $\tau_p^c$ to ensure that equation (\ref{otcorep}) is
satisfied for $z=1$.  Increasing $r$, the radial distance from the
accretor, has a similar effect on $\tau_p^c$ --- at larger radii the
viscous dissipation rate decreases, which leads to $\tau_p^c$ being
smaller for a given $\mdot_D$ (see Fig.~\ref{taufig}(c)).  Finally, 
Fig.~\ref{taufig}(d)
shows that the results are very similar for low-mass and high-mass
accreting black holes.  The slight difference between the two curves
in due to the fact that the mass of the central object determines the
physical scale ($H$) of the accretion flow (see discussion in section
\S\ref{ppbal} and Fig.~\ref{zsfig}(b)).

In studying pair-dominated corona solutions, two quantities of special
interest are the equilibrium temperature and the total optical depth
to electron scattering, $\tau_T = (1+2 z)\tau_p$, since they determine
to a large degree the shape of the emitted spectrum.  Fig.~\ref{hfig}
shows the dependence of these parameters on $\tau_p$ for different
values of the thin disk mass accretion rate.  On every curve, the
solution with $z=1$ (which corresponds to $\tau_p = \tau_p^c$) is
marked with a solid dot.  To the right of the dot, pairs are not
important.  Here the scattering optical depth of the corona is simply
equal to $\tau_p$, and the gas temperature increases with decreasing
optical depth, to keep up with the cooling.  In this regime, our
$\theta(\tau_p)$ curve is qualitatively similar to the results of HM,
except that for each value of $\mdot_D$, there is a maximum allowed
$\tau_p$, for which a corona can still be in thermal equilibrium.
Above this value, the bremsstrahlung and synchrotron emissivity of the
gas in the corona is larger than viscous heating, and thermal
equilibrium is not possible.

For $\tau_p < \tau_p^c$, on the other hand, the radiative transfer in
the corona is dominated by pairs.  In this regime, $\tau_T$ initially
{\it increases} with decreasing $\tau_p$, due to copious pair
production.  Correspondingly, $\theta$ decreases, since pairs more
than make up for dropping numbers of primary ionized electrons.  In
the limit $\tau_p \ll \tau_p^c$, the gas in the corona reaches Wien
equilibrium, where $z \propto 1/\tau_p$, and as a result, both
$\tau_T$ and $\theta$ converge to constant values, determined solely
by $\mdot_D$.  

A single temperature corona is gravitationally bound when the gas
temperature is below the virial value, which we obtain by equating the 
thermal energy and gravitational energy per proton in the gas: 
\begin{equation} 
\label{vir}
2 (1+z) \theta_{vir} = \frac{G M}{R c^2} \left[\frac{m_p}{m_e}+(1+2
z)\right].
\end{equation}
When $\theta > \theta_{vir}$ the gas in the corona simply escapes from
the system on the dynamical time scale.  Thus, the solutions
calculated here are consistent only for $\theta < \theta_{vir}$.  This
requirement is equivalent to restricting the scale height of the
corona (see equation (\ref{h})) to $h < 1/\beta$.  To check whether
this condition is satisfied, we have plotted in Fig.~\ref{hfig}(c)
$h(\tau_p)$ curves corresponding to equilibrium corona solutions. The
thin solid line is drawn for $h=2$, the stability limit for a corona
with equipartition magnetic fields ($\beta=0.5$).  Only those parts of
the curves that lie below this line correspond to gravitationally
bound solutions.  We have seen earlier that the solutions with $z>1$
exist only to the left of the dots.  In other words, pair-dominated
and stable coronae exist when $h<2$ and $\tau_p < \tau_p^c$.  These
two conditions are mutually exclusive for $\mdot_D < 10^{-6}$.

The dot-dashed curves in Fig.~\ref{hfig} show the properties of 
accretion disk coronae around a supermassive black hole, computed for
$mdot_D=10^{-2}$.  As we have seen already, the results are not very
sensitive to the black hole mass and there are no qualitative differences 
between $m=10$ and $m=10^8$.

\subsection{Two-Temperature Coronae} 
\label{rescor2} 
The gas in a two-temperature corona must satisfy both electron and
proton energy balance equations, which require that in thermal
equilibrium, compactness, viscous heating and Coulomb energy transfer
rate are all equal to one another.  First we concentrate on the balance
between the two latter quantities, which did not play any role in
single-temperature coronae.

The viscous dissipation rate is determined by the mass accretion rate
in the thin disk and $\delta$, and is independent of the amount of
matter in the corona.  The rate of proton cooling via Coulomb
collisions, however, varies as a product of the proton and
electron$+$positron densities in the corona, i.e. it is proportional
to $(1+2 z) \tau_p^2$.  On the low-$z$ branch, pair density is
negligible, so the Coulomb energy transfer rate scales simply as
$\tau_p^2$.  On the high-$z$ branch, we have $z \propto 1/\tau_p$, so
that this quantity is linear in $\tau_p$.  In either case, the protons
can be in thermal equilibrium only for relatively high values of
$\tau_p$.  On the other hand, in \S\ref{rescor1} we showed that
electron energy conservation in pair-dominated corona solutions
requires relatively low values of $\tau_p$.  It would appear that the
two conditions are mutually exclusive.

Of course in this simplistic argument we have neglected the dependence
of both proton cooling and heating rates on particle temperatures and
pair density.  To get the exact answer we have solved for
two-temperature corona solutions numerically, following the procedure
described in \S\ref{twotcor}, exploring the entire $\mdot_D - \tau_p$
parameter space.  The results are shown on Fig.~\ref{twotfig} for two
values of $\delta$.  Equilibrium solutions are allowed only in the
shaded region above the solid ($\delta=1.0$) or dashed ($\delta=0.1$)
line.  A comparison between Figs.~\ref{twotfig} and \ref{taufig}(b)
clearly shows that these regions lie far above the $\tau_p^c$ line,
and therefore, must have very few pairs.  And indeed, we find that in
every case, when an equilibrium corona solution is found, it
corresponds to $z < 1$.  In other words, we have been unable to find
equilibrium pair-dominated two-temperature corona solutions for any
choice of parameters.  

Our conclusions disagree with those of Kusunose \& Mineshige (1991), the
only authors, to our knowledge, who investigated $e^+ e^-$ pair
equilibrium states in a two-temperature corona.  This discrepancy stems
most likely from different assumptions made in the two studies about the
energy source in the corona.  In our model, the coronal gas is heated by
dissipating the gravitational energy of material in the disk, while
Kusunose \& Mineshige considered the situation where viscous heating is
due to differential rotation of the corona itself.  In a sense, their
model corresponds to an accretion flow with external soft photon
irradiation, rather than a canonical thin disk corona.  In fact, the
pair equilibrium solutions found by Kusunose \& Mineshige (1991) are
very similar to those discovered previously by Kusunose \& Takahara
(1988, 1990) and White \& Lightman (1989) in their study of an SLE
accretion disk.

\subsection{Hot Accretion Flows}
\label{resadaf}
Figs.~\ref{adaf1fig} and \ref{adaflfig} demonstrate how we solve for
equilibrium solutions in two-temperature hot accretion flows.  At a
fixed $f$, $\mdot$, and $r$, the scale height of the accreting gas is
fully determined, and therefore, we can compute $z$, $\ell$, and
$q_{ie} H^2 \sigma_T/(m_e c^2)$ (the dimensionless Coulomb energy
transfer rate) as functions of the electron temperature.  Compactness
and Coulomb energy transfer are plotted in Fig.~\ref{adaf1fig} as
solid$+$dotted and dashed$+$dot-dashed lines respectively.  As usual,
we have two branches for each value of $\theta_e$, one with negligible
pair density (dotted and dot-dashed lines), and the other dominated by
pairs (solid and dashed lines).  The transition between the two
branches occurs at $z=0.075$ on panel (a) and at $z=0.12$ on panel
(b).  The heating and cooling of the electrons (and pairs) in the flow
are balanced only at the intersection points marked in the figure.
The two panels show that we can find both high-$z$ and low-$z$
solutions to equation (\ref{adafe}), depending on the values of $f$
and $\mdot$.

It remains now to ensure that the proton energy balance equation
(\ref{adafp}) is satisfied.  Since the viscous dissipation rate has
the strongest dependence on the advection parameter $f$, in 
Fig.~\ref{adaflfig} we plot it (solid line) as a function of $f$ together
with the equilibrium proton cooling rate (dashed lines), obtained by
considering electron energy balance as discussed above.  The results
displayed on panels (a), (b), and (c) differ only by their values of
$\mdot$, listed in the figure caption.

It is clear from Fig.~\ref{adaflfig} that there are at most two values
of $f$ for which the gas is in thermal equilibrium.  The two equilibrium
points correspond to the well known solution branches for optically thin
accretion flows: the cooling-dominated branch ($f \ll 1$) discovered by
SLE, and the advection-dominated branch ($f \sim 1$) proposed by
Ichimaru (1977) and rediscovered by Narayan \& Yi (1994, 1995) and
Abramowicz et al. (1995).  Fig.~\ref{adaflfig}(c) shows that at high
accretion rates, both solutions disappear.  This feature of optically
thin accretion flow solutions was pointed out previously by many authors
(e.g. Narayan \& Yi 1995; Chen et al. 1995). 

The question we ask here is whether the properties of the optically
thin accretion flow solutions are modified because of the presence of
electron-positron pairs.  In Fig.~\ref{adaflfig}, the cooling rate
corresponding to the low-$z$ compactness branch (see Fig.~\ref{adaf1fig}) 
is indicated by long-dashed lines; short-dashed lines
show high-$z$ compactness curves.  The actual values of the pair
fraction $z$ corresponding to the compactness curves are plotted in
Fig.~\ref{adafzfig}.  As before, solid dots mark thermal equilibrium
solutions.  Note that in every case these equilibria have $z<1$,
implying that pairs do not play an important role in these accretion
flows.

To compare these results to previous work on pairs in optically thin
accretion flows, it is instructive to plot the thermal equilibrium
solutions for optically thin accretion flows at a fixed radius on a
$\log \mdot - \log \Sigma$ plane.  The resulting curves computed for
three different values of $\alpha$ are shown in Fig.~\ref{sigmafig}
(thick lines).  For comparison, thin lines show equilibrium solutions
computed assuming $z=0$.  Consistent with the findings of Kusunose \&
Mineshige (1996) and Bj\"ornsson et al. (1996), the effects of pairs
become more pronounced for larger values of $\alpha$; and are
practically negligible for $\alpha < 0.5$.  This is mainly because pair
production is more efficient in flows with higher density, and
therefore, mass accretion rate, and the critical $\mdot$ above which
thermal equilibrium solutions do not exist, scales roughly as
$\alpha^2$ (Narayan \& Yi 1995; Abramowicz et al. 1995).  However, even
for $\alpha=1$, pairs contribute no more than $\sim 30\%$ of the total
optical depth (see Fig.~\ref{sigmafig}).  In fact, we find that the
presence of magnetic fields in the flow reduces the importance of pairs
even further, as compared with pure bremsstrahlung solutions, since, for
a given $\alpha$, magnetized flows are restricted to lower values of
$\mdot$ due to extra cooling via synchrotron radiation. 

Though the results displayed in Fig.~\ref{sigmafig} were computed for $m=10$ we
find practically identical results for accretion flows around
supermassive black holes. 

\section{Discussion}
\label{disc}
From the analysis presented in this paper we conclude that
pair-dominated equilibrium solutions exist only in single-temperature
disk--plus--corona models.  Both two-temperature coronae and
single-phase hot accretion flow models have pair fractions $z=n_+/n_p$
significantly below unity.  What is the special feature of
single-temperature coronae that allows them to be pair-dominated? The
answer seems to be that it has fewer constraints then the other two
models.

The gas in the corona is heated via viscous dissipation of energy
in an underlying thin disk.  The rate of such heating is
completely determined by the properties of the thin disk (namely, mass
accretion rate $\mdot_D$ and radius $r$) and is virtually independent
of the amount or temperature of matter in the corona, which specify
the cooling rate.  Thus, for a given value of $\mdot_D$, the energy
balance in the corona allows us to solve for either $\tau_p$ or
$\theta$, but not both.  One of these parameters has to be specified
as an initial condition.  This extra degree of freedom allows us to
adjust the mass accretion rate in the disk and the optical depth of
the corona to obtain thermal equilibrium solutions with $z>1$.  We
find that such solutions are restricted to $\tau_p < \tau_p^c$, where
the critical value of the proton optical depth decreases with
decreasing $\mdot_D$ (see Fig.~\ref{taufig}).

In a two-temperature corona, heating and cooling rates are still
relatively independent of each other; however, proton energy balance
imposes an important additional constraint on the model.  The proton
cooling rate via Coulomb scattering between protons and electrons (as
well as pairs) is sensitive to the number density of particles in the
corona, while proton heating through viscous dissipation is
independent of $\tau_p$.  Consequently, protons are in thermal balance
only for relatively high values of $\tau_p$.  But at these values of
$\tau_p$, electron energy balance requires $z <1$ (Fig.~\ref{twotfig}).

In the single-phase hot accretion flow model, both viscous heating and
radiative cooling rates are determined by the mass accretion rate in
the flow.  In this model, $\mdot$ specifies both the optical depth and
temperature of the accreting gas; one can no longer adjust these
variables independently in order to obtain pair-dominated solutions.
We find that for a realistic range of values for the viscosity
parameter $\alpha$, the pair density in hot accretion flows is
negligible, both for the cooling-dominated (SLE) and the
advection-dominated (Ichimaru 1977, Narayan \& Yi 1994) solution
branches.  In the two-temperature accretion flow solution we have
considered, the Coulomb energy transfer bottleneck also plays a role
in limiting the effects of pairs.

Even in one-temperature coronae, we find that pair-dominated solutions
exist only over a relatively narrow range of the proton optical depth,
for each value of $\mdot_D$ (see Fig.~\ref{hfig}(c)).  It is bounded from above
by $\tau_p^c$, imposed by the thermal equilibrium balance equation for
the gas, and from below by the requirement that the hot gas in the
corona should be gravitationally bound.  The latter condition can be
somewhat relaxed, if we assume that the corona is held in place not by
gravity, but rather by magnetic field loops anchored in the thin disk.
In fact, the most popular recent scenario for the dissipation of the
thin disk gravitational energy in the corona is based on reconnection of
the magnetic field loops generated in the disk and transported by
buoyancy into the corona (Galeev, Rosner \& Vaiana 1979; Field \&
Rogers 1993).  In this picture, if the magnetic energy density dominates
the energy density of the gas, magnetic fields may prevent the escape of
particles from the corona even if their temperature is above virial.  On
the other hand, if the cooling of the gas in the corona is dominated by
inverse Compton scattering of the thin disk photons, electrons and
positrons can undergo a considerable radiative acceleration along the
accretion disk poles (see for example Liang \& Li [1995], Li \& Liang
[1996]), which may allow the escape of pairs with temperatures below
virial. 

When we consider a one-temperature accretion disk corona with
$\theta_e = \theta_p$, an important question to ask is what keeps
electrons and protons at the same temperature.  If viscous dissipation
preferentially heats the protons, as is often assumed in the
literature (e.g. SLE; Rees et al.  1982; Narayan \& Yi 1995), the
electrons will be considerably cooler than the heavier particles.  Our
results for two-temperature coronae clearly show that Coulomb energy
transfer is too inefficient to equilibrate $\theta_e$ and $\theta_p$.
One way to solve this problem is to introduce some other more
efficient energy transfer mechanism between the protons and electrons
in the corona (e.g. Begelman \& Chiueh 1988).  On the other hand,
recent work on dissipation mechanisms for MHD waves by Quataert (1998)
and Gruzinov (1998) indicate that in magnetically dominated plasma it
is electrons, rather than the protons that are preferentially heated.
In this case, the protons might even be cooler than the electrons (if
the accretion timescale is shorter than the Coulomb energy transfer
timescale).  However, since it is the latter that contribute to
cooling and pair-production, $\theta_p$ is practically irrelevant, so
that the corona can be treated as a single-temperature gas cloud.

Finally, we would like to emphasize that the results for thin disk
coronae presented in this paper apply only to steady-state,
homogeneous corona scenarios such as discussed, for example, by Haardt
\& Maraschi (1991, 1993), as opposed to time-dependent flare models in
which a corona contains many localized active regions (e.g.  Haardt,
Maraschi \& Ghisellini 1994; Stern et al. 1995b, Nayakshin \& Melia
1997a,b; Di Matteo, Celotti \& Fabian 1997, 1998).  In the latter case,
many of the constraints imposed to find the equilibrium solutions need
not be satisfied.  In particular, the protons may transfer only part
of the viscously dissipated energy to the electrons and pairs.  In
that case, the protons will experience a net heating, until they reach
virial temperature and escape from the system, perhaps in the form of
an outflow or a jet.  After that more material will be evaporated into
the corona from the disk, and the cycle will be repeated.  Similarly,
the pairs themselves may be periodically ejected from the corona
driven either by radiative acceleration (Liang \& Li 1995) or simply
through heating above their virial temperature, in the absence of
energy balance.  Such transient or time-dependent situations are
beyond the scope of the paper.

\section{Summary}
\label{sum}
We find that the properties of pair equilibria in static plasma
clouds, taking into account synchrotron emission and Comptonization of
external soft photons differ only in quantitative details from the
results of S84, who restricted his study to Comptonized
bremsstrahlung and double Compton emission.  For every value of the
proton optical depth, $\tau_p$, there exists a critical electron
temperature $\theta_{max}(\tau_p)$ below which there are two pair
equilibrium solution branches, one of which is pair-dominated, and the
other practically pair-free.  On the pair-free branch,
particle-particle pair production processes dominate and the pair
fraction $z = n_+/n_p$ is independent of $\tau_p$.  By contrast, on
the pair-dominated branch the electron (and positron) scattering
optical depth is constant, so that $z \propto 1/\tau_p$.  The extra
sources of photons considered here (synchrotron and external photons)
do not change the general character of the solutions, but they do
decrease the value of $\theta_{max}$ so that pair equilibria are
restricted to lower temperatures than in S84.
 
We studied in this paper three popular accretion scenarios involving
hot plasmas to assess the importance of pairs.  Our results for the
three models are summarized below:
 
\begin{itemize}  
\item[1)] In the case of a single-temperature corona above a thin
disk, the cooling of the gas must be balanced by viscous heating.  The
latter quantity is determined by the mass accretion rate, $\Mdot_D$,
in the thin disk and the fraction $\delta$ of the viscous energy which
is dissipated in the corona.  The compactness on the low-$z$ branch is
due primarily to inverse Compton scattering of synchrotron and thin
disk photons, so that $\ell \propto \tau_p$.  As a result, for every
value of the mass accretion rate in the thin disk, we find that there
is a critical $\tau_p^c$ in the corona below which the equilibrium
solution is pair-dominated (see Fig.~\ref{taufig}).  For larger values
of $\tau_p$, pairs in the corona are unimportant. Thus, while pure
pair coronae are allowed, such coronae have very little mass and are
dominated by energy deposition from the disk.
 
\item[2)] A two-temperature corona above a thin disk must satisfy two
energy conservation equations, one for electrons and one for protons.
That is, the Coulomb energy transfer rate must balance both the rate
of viscous energy input into the protons and the the cooling rate of
the electrons.  Since the Coulomb energy transfer is a two-body
process, it is proportional to $\tau_p^2$ on the low-$z$ pair
equilibrium branch and goes as $z \tau_p^2 \propto \tau_p$ on the
high-$z$ branch.  We find that the extra constraint on the energy
balance (compared to a single-temperature corona) restricts solutions
to relatively high values of the proton optical depth.  Only at such
$\tau_p$ is Coulomb coupling efficient enough.  However, the critical
$\tau_p$ is generally above the maximum $\tau_p^c$ up to which
pair-dominated coronae are possible (see Figs.~\ref{taufig} and
\ref{twotfig}).  Thus, we conclude that, if viscous dissipation heats
primarily the protons in the corona, the equilibrium solutions are
generally pair-free.
 
\item[3)] Finally, we have considered pair equilibria in hot accretion flows
described by the self-similar solution of Narayan \& Yi (1995).  We find
that the energy transfer bottleneck due to Coulomb coupling conspires
with the advective cooling in these flows to limit the effects of $e^+
e^-$ pairs.  We agree with the results of earlier investigators
(Bj\"ornsson et al.  1996 and Kusunose \& Mineshige 1996) who concluded
that for reasonable values of the viscosity parameter ($\alpha \lsim
0.5$) there are no pair-dominated solutions on either the
advection-dominated or cooling-dominated (i.e. SLE) branches.  Previous
work dealt only with Comptonized bremsstrahlung cooling.  We have included
Comptonized synchrotron here, but recover the same results as before
with only minor quantitative differences (Fig.~\ref{sigmafig}).
\end{itemize}

\acknowledgements

I would like to thank Ramesh Narayan for numerous useful discussions
and suggestions. This work was supported in part by National Science
Foundation Graduate Research Fellowship, and by NASA through AXAF
Fellowship grant \#PF8-10002 awarded by the AXAF Science Center, which
is operated by the Smithsonian Asprophysical Observatory for NASA
under contract NAS8-39073.

\newpage
\begin{appendix}
\section{Inverse Compton Scattering}
\label{apa}

\subsection{Analytical Formalism}
We treat inverse Compton scattering of photons by thermal electrons
using the approximate analytical formulae derived by S84
and Esin et al. (1996).  

When the plasma cloud is optically thin, $\tau_T < 1$, the probability
$\cP_j$ that a photon will suffer exactly $j$ scatterings before
escaping can be approximated by a Poisson formula:
\begin{equation}
\label{p}
\cP_j = \frac{e^{-s} s^j}{j!},
\end{equation}
where $s = \tau_T + \tau_T^2$ is the mean number of scatterings
(Rybicki \& Lightman 1979).  If the initial photon energy, $x = h
\nu/m_e c^2$ is small compared with the thermal energy of the
electrons in the cloud, on average, each scattering will increase the
photon energy by a factor $A = 1 + 4\theta_e + 16 \theta_e^2$ (Rybicki
\& Lightman 1979).  This process saturates when the photon energy
becomes comparable to the average energy of the Wien distribution, $3
\theta_e$.  Thus, the number of scatterings corresponding to
saturation is $j_{m} = \ln{(3 \theta_e/x)}/\ln{A}$.  Then the fraction
of photons with initial energy $x$ that are scattered into the Wien
peak before escaping is simply
\begin{equation}
\label{f}
f(x) = \sum_{j=j_m}^{\infty}{\cP_j} = \frac{3 \theta_e}{x} P(j_m,s),
\end{equation}
where $P(j_m,s)=\int_0^s{e^{-y} y^{j_m-1} d y}/\Gamma{(j_m)}$, is the
incomplete gamma function (see Esin et al. [1996] for a detailed
derivation of this formula).  

The energy gain through Comptonization is characterized by the average
energy enhancement factor for each photon, $\xi$, defined as the ratio
of the photon final and initial energies:
\begin{equation}
\label{xi}
\xi(x) = \sum_{j=0}^{j_m}{A^j \cP_j} + A^{j_m}
\sum_{j=j_m+1}^{\infty}{\cP_j} = e^{(A-1) s} [1-P(j_m+1,A s)] + \frac{3
\theta_e}{x} P (j_m+1,s).
\end{equation}

For an optically thick cloud, $\tau_T > 1$, we use a different
expression for probability $\cP_j$, derived by Sunyaev \& Titarchuk
(1980, see also S84),
\begin{equation}
\label{pp}
\cP_j = \frac{e^{-j/(3 \tau_T^2)}}{3 \tau_T^2} .
\end{equation}
Since in this regime each photon will undergo many scatterings before
escaping, we can treat $j$ as a continuous variable (the expression
for $\cP_j$ is normalized accordingly).  Then the fraction of photons
scattered into the Wien peak can be written as an integral:
\begin{equation}
\label{fp}
f(x) = \int_{j=j_m}^{\infty}{\cP_j} =
\exp{\left[-\frac{\ln{(3 \theta_e/x)}} {3 \tau_T^2 \ln{A}}\right]}.
\end{equation}
Note that this expression differs slightly from that derived by S84
(equation C3), since we assume that the average energy of Wien photons
is $3 \theta_e$, rather than $\theta_e$. 

The average energy enhancement factor for photons is then 
\begin{equation}
\label{xip}
\xi(x) = \int_{0}^{j_m}{A^j \cP_j} + A^{j_m}
\int_{j_m}^{\infty}{\cP_j} = \left[\frac{(3 \theta_e/x) 
f(x) -1}{3 \tau_T^2 \ln{A} -1}\right] + \frac{3 \theta_e}{x}
f (x).
\end{equation}

In our calculations we simply use equations (\ref{f}) and (\ref{xi})
when $\tau_T < 1$ and equations (\ref{fp}) and (\ref{xip}) when $\tau_T
> 1$, with a suitable interpolation across $\tau_T = 1$.

\subsection{Comptonization of Bremsstrahlung Photons}

Bremsstrahlung photon emissivity is given by a normalized expression (S84)
\begin{equation}
S_B dx = 2 \frac{\ln{(\theta_e/x)}}{\ln^2{(\theta_e/x_B)}} \frac{d
x}{x},\ \ \ \ x_m<x<\theta_e,
\end{equation}
where $x_B$ is the bremsstrahlung self-absorption energy, computed by
comparing the bremsstrahlung and Compton scattering absorption
coefficients (for details see appendix D of S84).  Then the
emission weighted  fraction $f_B$ is simply 
\begin{equation}
\label{fb}
f_B = \int_{x_B}^{\theta} {f(x) S_B(x) d x},
\end{equation} 
while the total energy amplification factor for bremsstrahlung emission
is
\begin{equation}
\label{xib}
\xi_B = \frac{\int_{x_B}^{\theta} {x \xi(x) S_B(x) d x}}{\int_{x_B}^{\theta}
{x S_B(x) d x}},
\end{equation} 

In the optically thick limit, equations (\ref{fb}) and (\ref{xib}) can
be evaluated analytically (S84):
\begin{equation}
f_B = 2 \exp{\left(-\frac{\ln{3}}{3 \tau_T^2 \ln{A}}\right)} [y^2 -
(y+y^2) e^{-1/y}],
\end{equation}
and
\begin{equation}
\xi_B = 1 + f_B \frac{3}{4} \ln^2{\left(\frac{\theta_e}{x_B}\right)}.
\end{equation}
where $y= 3\tau_T^2 \ln{A}/\ln{(\theta_e/x_B)}$ is the emission
averaged Compton $y$-parameter.

In the optically thin regime, we integrate equations
(\ref{fb}) and (\ref{xib}) numerically.  

The rate of Wien photon production due to Compton upscattering of
bremsstrahlung photons is simply $f_B \dot n_{\gamma}^B$.  We use the
expression for $n_{\gamma}^B$, volume emissivity of bremsstrahlung
photons via $e^{\pm} e^{\pm}$ and $e^{\pm} p$ interactions, derived by
S84 (equations A1-A3 and A22a).  Total Comptonized bremsstrahlung
cooling rate per unit volume is $\xi_B q_B$, where $q_B$ is the
thermal bremsstrahlung cooling rate per unit volume (Svensson 1982). 

\subsection{Comptonization of Synchrotron Photons}

Thermal synchrotron emission is generally strongly self-absorbed, so
that most of the emission comes out near the self-absorption frequency
$x_S$, calculated as described in Esin et al. (1996).  When calculating
the effect of Comptonization, this means we can treat synchrotron
emission as a monochromatic source of photons, characterized by a
photon production rate $\dot n_{\gamma}^S = q_S/(x_S m_e c^2)$.  Here $q_S$
is the synchrotron cooling rate per unit volume, for which we employ
the expressions derived by Mahadevan et al. (1995) and Esin et
al. (1996).  Then the rate of Wien photon production due to
upscattering of synchrotron photons is $f_S \dot n_{\gamma}^S = f(x_S)
\dot n_{\gamma}^S$.  Similarly, total Comptonized synchrotron cooling rate
is just $\xi_S q_S = \xi(x_S) q_S$.

\subsection{Comptonization of Thin Disk Photons}

Since thin disk emission has a thermal spectrum, we can treat it in
exactly the same way as the synchrotron emission -- as a monochromatic
source of photons with energy $x_D = 2.8 k T_D/m_e c^2$.  The
characteristic temperature $T_D$ is given by equation (\ref{td}).
Then the rate of production of the thin disk photons per unit volume
of the corona is $\dot n_{\gamma}^D = Q_D/(H x_D m_e c^2)$, where
cooling rate per unit area of the disk, $Q_D$, is given by equation
(\ref{disken}).  Consequently, we can write the rate of upscattering of
disk photons into the Wien peak as $f_D \dot n_{\gamma}^D = f(x_D)
\dot n_{\gamma}^D$.

The cooling rate of gas in the corona through inverse Compton
scattering of thin disk photons is then $(\xi_D-1) Q_D/H = 
(\xi(x_D)-1) Q_D/H$.  

\section{Emission from the Thin Disk}
\label{apb}

Following HM we approximate disk and corona as
two uniform adjacent slabs, characterized by different temperatures
and densities.  The rate of gravitational energy release per unit area
is $Q_G$, of which a fraction $\delta$ is dissipated directly in the
corona, while the remaining $(1-\delta) Q_G$ is dissipated within the
thin disk.  

The cold optically thick disk radiates as a blackbody, with a total
flux $Q_D$.  On the other hand, the optically thin corona cools by
emitting Comptonized synchrotron and bremsstrahlung radiation, as well
as through inverse Compton scattering of the thin disk photons.  Total
cooling rate per unit ares of the corona can be written as $Q_C = Q_{SB} +
(\xi-1) Q_D$, where the first term stands for the internal coronal
emission and $\xi$ is amplification factor due to Comptonization of
the thin disk emission.  The energy balance for the corona requires that
\begin{equation}
\label{corbal}
\delta Q_G = Q_{SB} + (\xi-1) Q_D.
\end{equation}

We assume further that a fraction $\eta$ of all photons emitted of
scattered in the corona is directed downwards towards the thin disk,
while the remaining radiation escapes towards the observer.  Then the
total flux incident onto the thin disk can be written as 
\begin{equation}
\label{qinc}
Q_{inc} = \eta [Q_{SB} + (\xi-1) Q_D + Q_D (1-\exp{(-\tau_T)})],
\end{equation}
where $\tau_T$ is the optical depth of the corona to electron
scattering, and $(1-\exp{(-\tau_T)})$ is the fraction of the thin disk
photons scattered at least once before escaping.  This incident
radiation is partly absorbed by the thin disk, heating it further, and
partly reflected.  We write the absorbed fraction as $(1-a) Q_{inc}$.

Note that equation (\ref{qinc}) above differs from the corresponding
expression in HM in two ways.  Firstly, we include the internal
emission from the corona.  Secondly, we take into account the fact
that the corona not only amplifies thin disk emission through Compton
scattering, it also acts as a reflector for the thin disk photons.  In
the limit when Comptonization is not important, i.e. $\xi \simeq 1$,
the presence of the corona still raises the equilibrium temperature of
the disk through the greenhouse effect.

With this, we can write down the energy balance equation for the disk:
\begin{equation}
\label{diskbal}
(1-\delta Q_G) + (1-a) \eta [Q_{SB} + (\xi-1) Q_D + Q_D (1-\exp{(-\tau_T)})] = 
Q_D.
\end{equation}
Combining equations (\ref{corbal}) and (\ref{diskbal}), we obtain the
expression for total disk emission, which is independent of the
details of inverse Compton scattering in the corona:
\begin{equation}
Q_D = Q_G \frac{1 - \delta + \delta \eta (1-a)}{1-\eta (1-a) 
(1-\exp{(-\tau_T)})}.
\end{equation}
\end{appendix}

\vfill\eject 
\references 
\def\refpar{\hangindent=3em\hangafter=1}
\def\reference{\refpar\noindent} 
\def\apj{ApJ} 
\def\apjs{ApJS}
\def\mnras{MNRAS} 
\def\aa{A\&A} 
\def\aas{A\&A Suppl. Ser.}
\def\aj{AJ} 
\def\araa{ARA\&A} 
\def\nat{Nature} 
\def\pasj{PASJ}

\reference Abramowicz, M. A., Chen, X., Kato, S., Lasota, J. P., \& Regev, O. 1995, \apj, 438, L37

\reference Begelman, M. C. \& Chiueh, T. 1988,\apj, 332, 872

\reference Bj\"ornsson, G., Abramowicz, M. A., Chen, X., \& Lasota, J.-P. 1996, \apj, 467, 99  
 
\reference Bj\"ornsson, G. \& Svensson, R. 1991a, \mnras, 249, 177
 
\reference Bj\"ornsson, G. \& Svensson, R. 1991b, \apj, 371, L69
 
\reference Bj\"ornsson, G. \& Svensson, R. 1992, \apj, 394, 500 

\reference Chen, X. 1995, \mnras, 275, 641
 
\reference Chen, X., Abramowicz, M. A., Lasota, J. P., Narayan, R., Yi, I. 1995, \apj, 443, L61

\reference Di Matteo, T., Blackman, E. G., \& Fabian, A. C. 1997, \mnras, 291, L23

\reference Di Matteo, T., Celotti, A., Fabian, A. C. 1997, \mnras, 291, 805

\reference Di Matteo, T., Celotti, A., Fabian, A. C. 1998, astro-ph/9805345
 
\reference Esin, A. A., Narayan, R., Ostriker, E., \& Yi, I. 1996, \apj, 465, 312
 
\reference Field, G. B., \& Rogers, R. D. 1993, \apj, 403, 94 

\reference Frank, J., King, A., \& Raine, D. 1992, Accretion Power in Astrophysics (Cambridge, UK: Cambridge University press)

\reference Galeev, A. A., Rosner, R. \& Vaiana, G. S. 1979, \apj, 229, 318 

\reference Gruzinov, A. V. 1998, \apj, in press

\reference Haardt, F. \& Maraschi, L. 1991, \apj, 380, 51 (HM)

\reference Haardt, F. \& Maraschi, L. 1993, \apj, 413, 507

\reference Haardt, F., Maraschi, L. \& Ghisellini, G. 1994, 432, L92 

\reference Nayakshin, S. \& Melia, F. 1997a, \apj, 490, L13

\reference Nayakshin, S. \& Melia, F. 1997b, submitted to {\apj}L, astro-ph/9709286

\reference Ichimaru, S. 1977, \apj, 214, 840

\reference Li, H. \& Liang, E. P. 1996, \apj, 458, 514

\reference Liang, E. P. 1991, \apj, 367, 470

\reference Liang, E. P. \& Li, H. 1995, \aa, 298, L45

\reference Lightman, A. P. 1982, \apj, 253, 842
 
\reference Kusunose, M. 1987, \apj, 321, 186
 
\reference Kusunose, M. \& Mineshige, S. 1996, \apj, 468, 330
 
\reference Kusunose, M. \& Mineshige, S. 1991, \apj, 381, 490
 
\reference Kusunose, M. \& Takahara, F. 1988, \pasj, 40, 435 
 
\reference Kusunose, M. \& Takahara, F. 1989, \pasj, 41, 263
 
\reference Kusunose, M. \& Takahara, F. 1990, \pasj, 42, 347
 
\reference Kusunose, M. \& Zdziarski, A. A. 1994, \apj, 422, 737
 
\reference Mahadevan, R., Narayan, R., Yi, I. 1996, \apj, 465, 327

\reference Nakamura, K. E., Kusunose, M., Matsumoto, R., Kato, S. 1997, \pasj, 49, 503
 
\reference Narayan, R. \& Yi, I. 1994, \apj, 428, L13

\reference Narayan, R. \& Yi, I. 1995, \apj, 452, 710
 
\reference Poutanen, J., Krolik, J. H., \& Ryde F. 1997, \mnras, 292, L21

\reference Quataert, E. 1998, \apj, in press

\reference Rees, M. J., Begelman, M. C., Blandford, R. D., Phinney, E. S. 
1982, \nat, 295, 17

\reference Rybicki, G. B. \& Lightman, A. P. 1979, Radiative Processes in Astrophysics (New York: John Wiley \& Sons)

\reference Shakura, N. I. \& Sunyaev, R. A. 1973, \aa, 24, 337

\reference Shapiro, S. L., Lightman, A. P., \& Eardley, D. M. 1976, \apj, 204, 187 (SLE)

\reference Sikora, M. \& Zbyszewska, M. 1986, Acta Astron., 36, 255
 
\reference Skibo, J. G., Dermer, C. D., Ramaty, R., McKinley, J. M. 1995, \apj, 446, 86
 
\reference Stepney, S. \& Guilbert, P. W. 1983, \mnras, 204, 1269

\reference Stern, B. E., Begelman, M. C., Sikora, M., \& Svensson, R. 1995a, \mnras, 272, 291
 
\reference Stern, B. E., Poutanen, J., Svensson, R., Sikora, M., Begelman, M. C. 1995b, \apj, 449, L13 
 
\reference Svensson, R. 1982, \apj, 258, 335 
 
\reference Svensson, R. 1984, \mnras, 209, 175 (S84)
 
\reference Tritz, B. G. \& Tsuruta, S. 1989, \apj, 340, 203
 
\reference White, T. R. \& Lightman, A. P. 1989, \apj, 340, 1024
 
\reference White, T. R. \& Lightman, A. P. 1990, \apj, 352, 495 
 
\reference Zdziarski, A. A. 1986, \apj, 303, 94
 
\reference Zdziarski, A. A. et al. 1994, \mnras, 269, L55

\vfill\eject

\setcounter{figure}{0}
\begin{figure}
\includegraphics{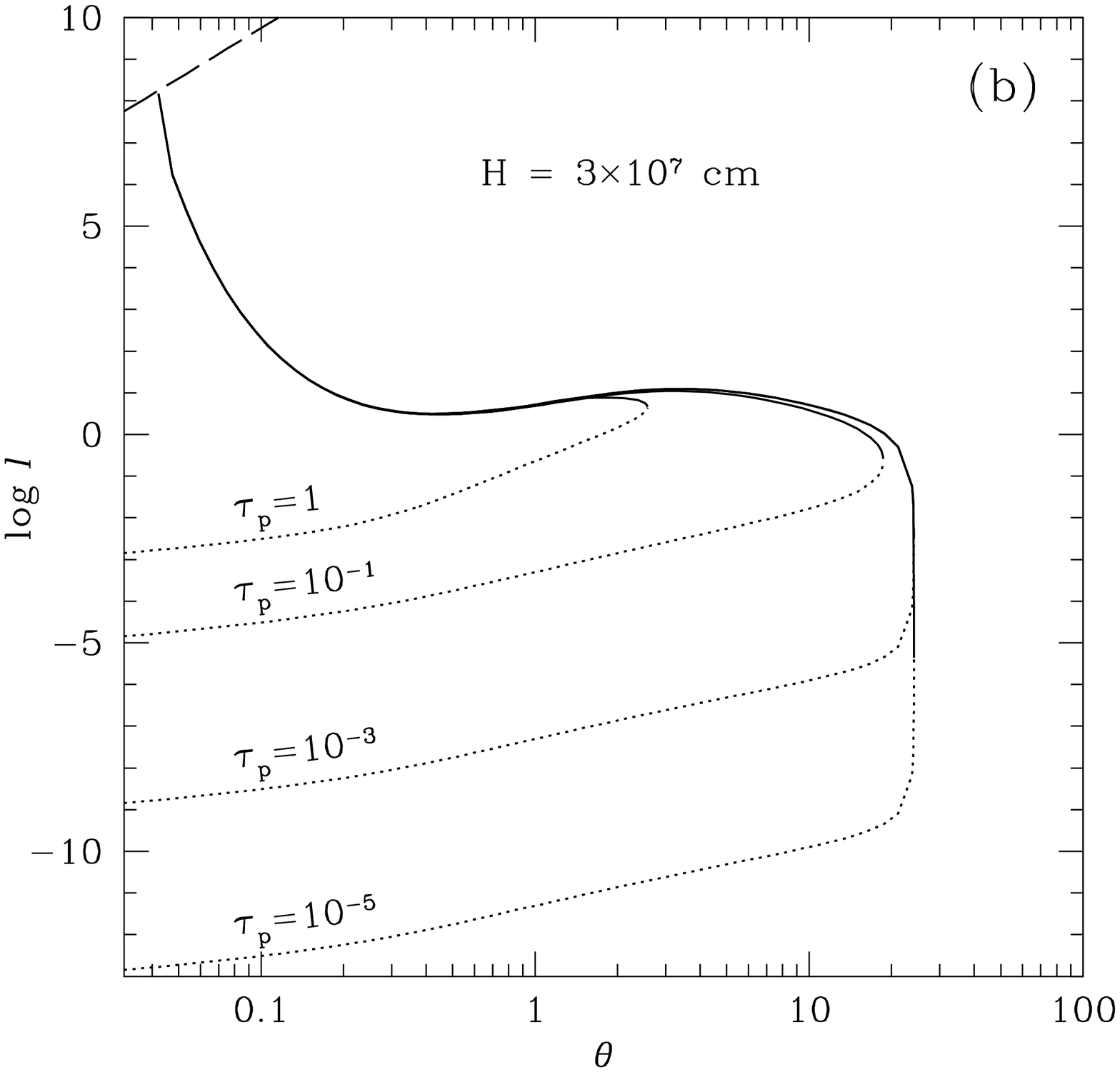} 
\vskip 6.5in 
\caption{\label{lfig} (b) Compactness as a function of plasma
temperature corresponding to the equilibrium pair fraction solutions
in panel (a).  Solid lines correspond to the high-$z$ branch and
dotted lines to the low-$z$ branch.  The blackbody limit is
indicated by a long-dashed line.}
\end{figure}

\begin{figure}
\includegraphics{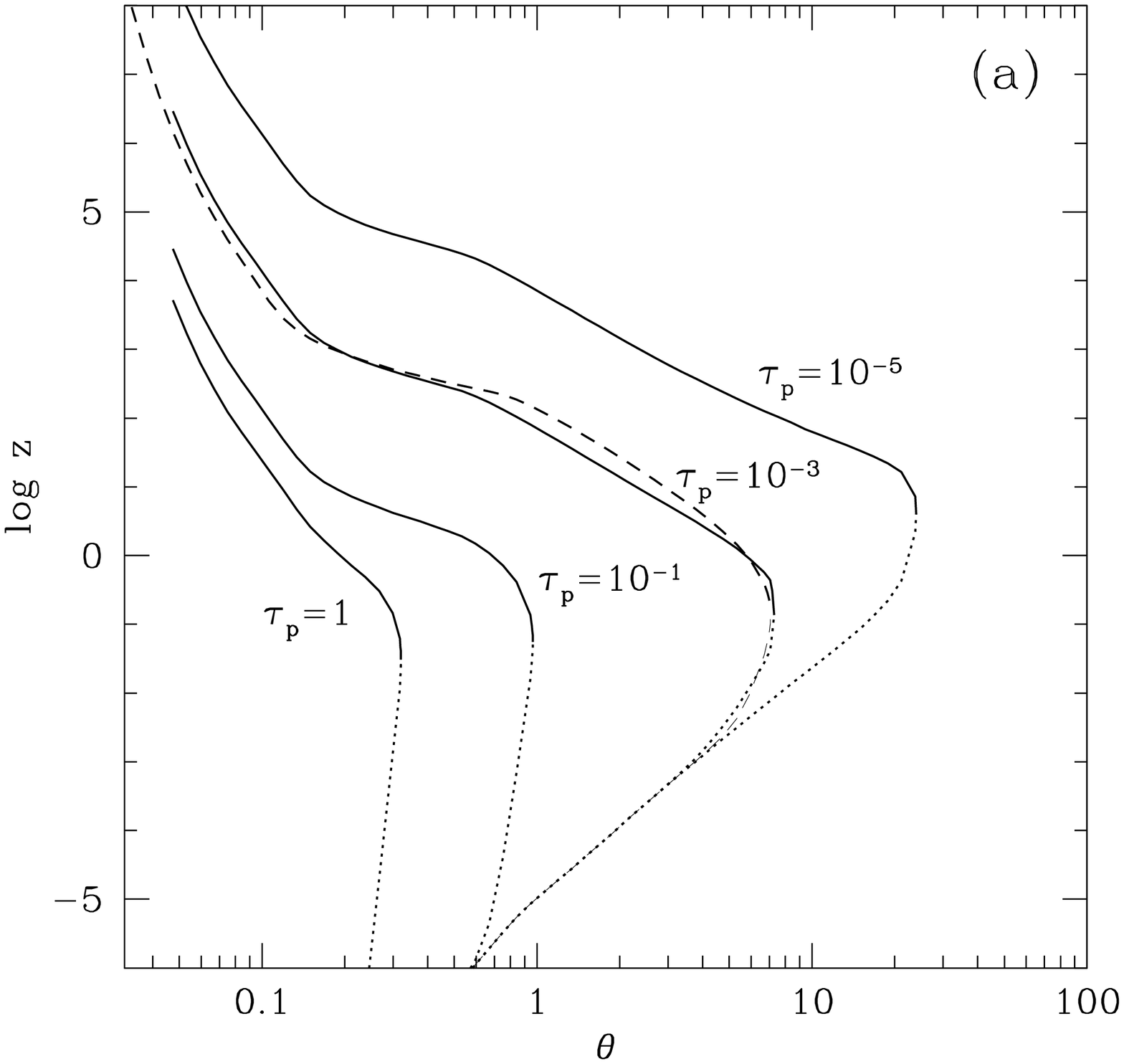} 
\vskip 6.5in 
\caption{\label{zsfig} (a) The equilibrium pair fraction, $z = n_+/n_p$, in a
one-temperature plasma cloud cooling via Comptonized bremsstrahlung
and synchrotron, shown as a function of gas temperature.  In
calculating the synchrotron emissivity we assumed that the magnetic
pressure is equal to the gas pressure.  As in Fig.~1, solid and dotted
lines show the high-$z$ and low-$z$ branches, respectively.  These four
curves were computed for a cloud of size $H = 3\times 10^7\,{\rm cm}$,
but with different values of the proton optical depth, as shown on the
figure.  The dashed curve shows the pair fraction computed for
$\tau_p=10^{-3}$ and $H = 3\times 10^{14}\,{\rm cm}$.}
\end{figure}

\setcounter{figure}{1}

\begin{figure}
\includegraphics{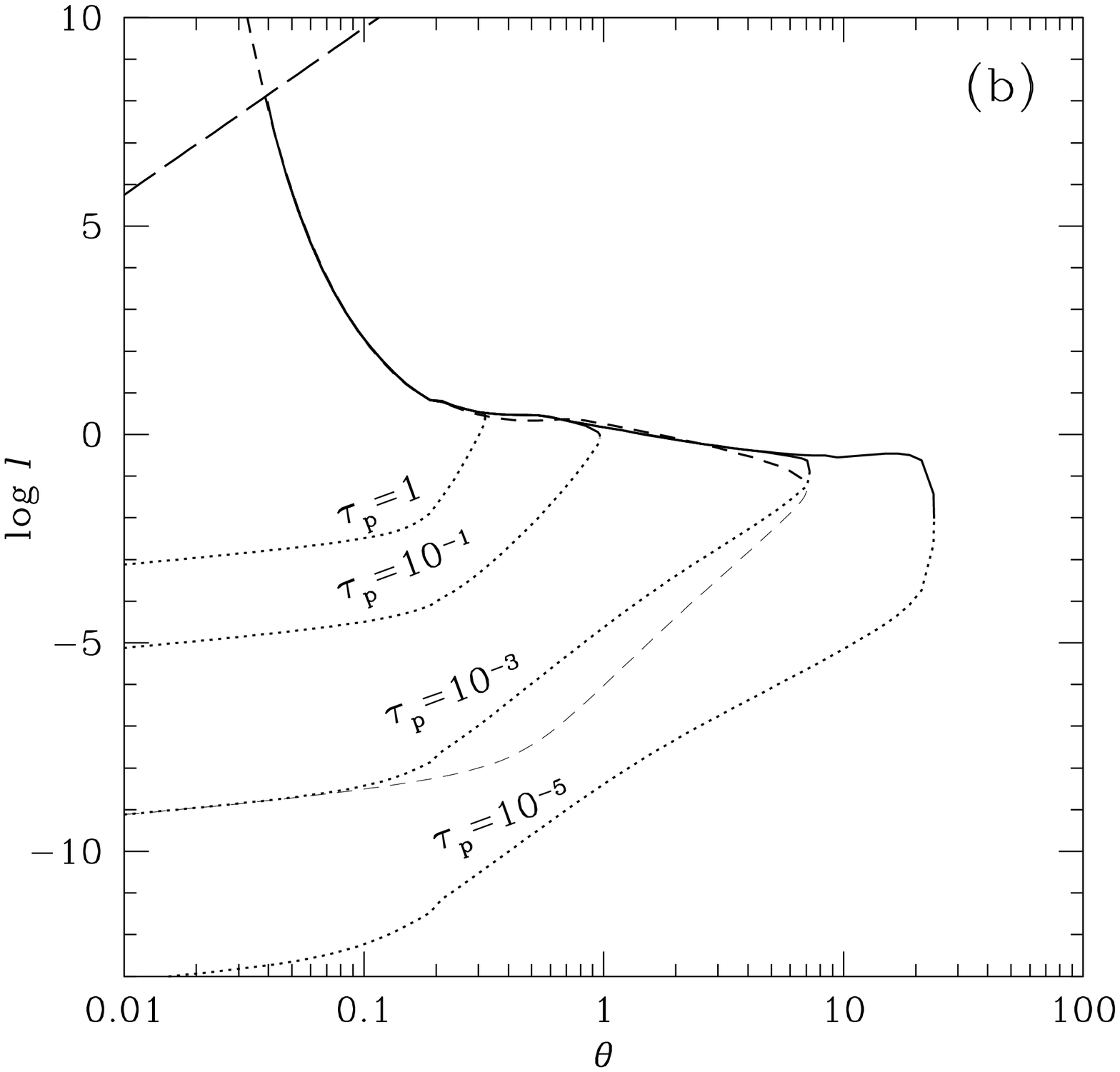} 
\vskip 6.5in 
\caption{\label{lsfig} (b) Compactness as a function of plasma temperature
corresponding to the equilibrium pair fraction solutions in Figure
2(a).  The blackbody limit for a cloud of size $H = 3\times 10^7\,{\rm
cm}$ is indicated by a long-dashed line.}
\end{figure}

\begin{figure}
\includegraphics{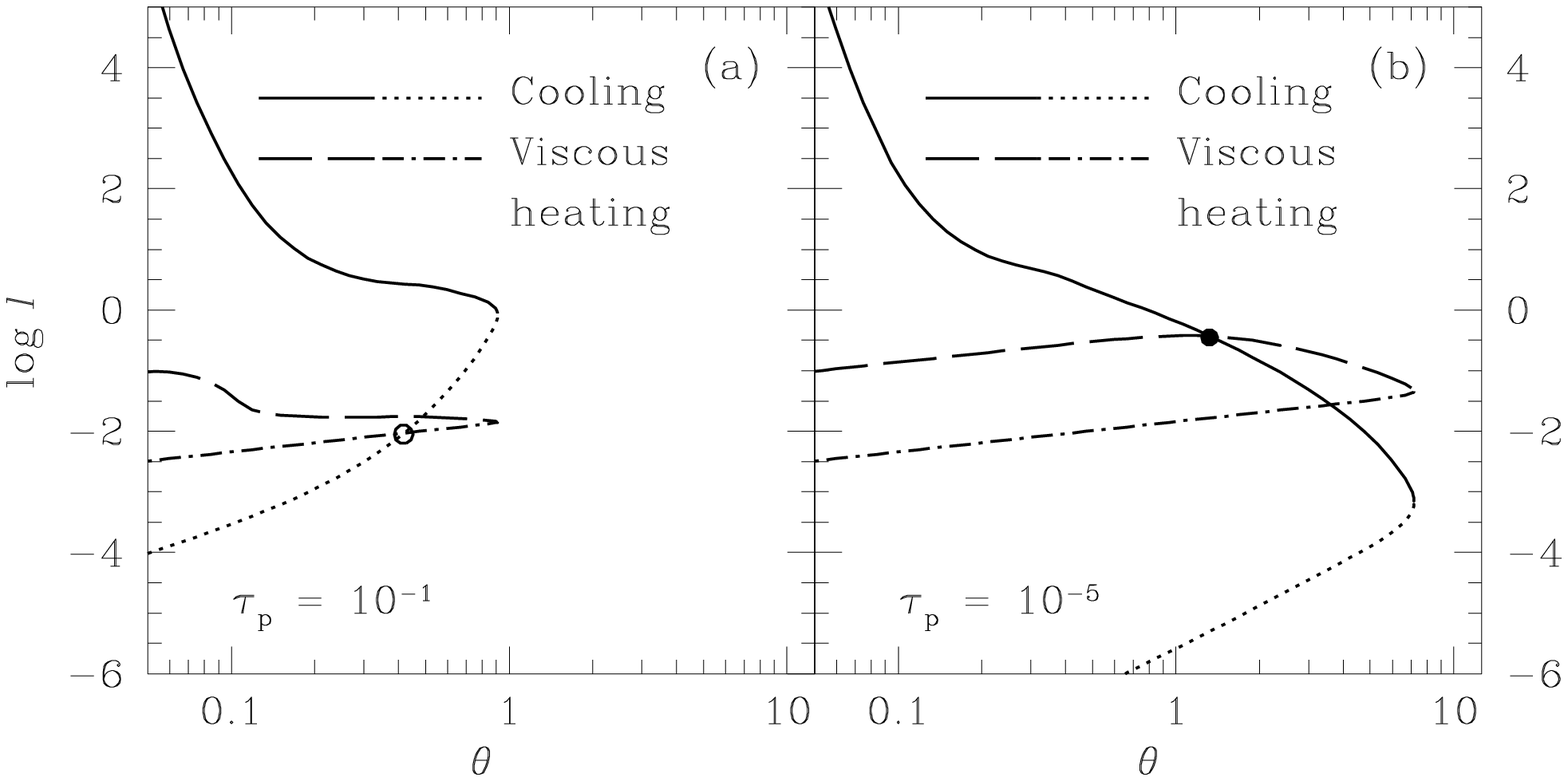} 
\vskip 5.5in 
\caption{\label{corfig} Illustrates the balance between dimentionless
viscous heating, $\delta Q_G H \sigma_T/(m_e c^2)$, and cooling,
$\ell$, rates in a magnetized corona above a thin disk with $\mdot_D =
10^{-3}$, $r=10$, $m=10$.  The values of compactness and viscous
dissipation on the high-$z$ solution branch are shown as solid and
dashed lines respectively, and the corresponding quantities for the
low-$z$ solution values are shown as dotted and dot-dashed lines.  (a)
Shows the case of a corona with $\tau_p= 0.1$.  There is a low-$z$
equilibrium solution, marked with an open circle.  This solution is
practically pair-free with $z=2\times 10^{-5}$.  Note that there is no
high-$z$ solution for this value of $\tau_p$, since the solid and
dashed curves do not intersect each other. (b) Shows high-$z$
equilibrium solution for a corona with $\tau_p=10^{-5}$, marked with a
solid dot.  The solutions corresponds to a pair fraction $z=1.1\times
10^{3}$.  For this value of $\tau_p$ there is no low-$z$ solution.}
\end{figure} 

\begin{figure}
\includegraphics{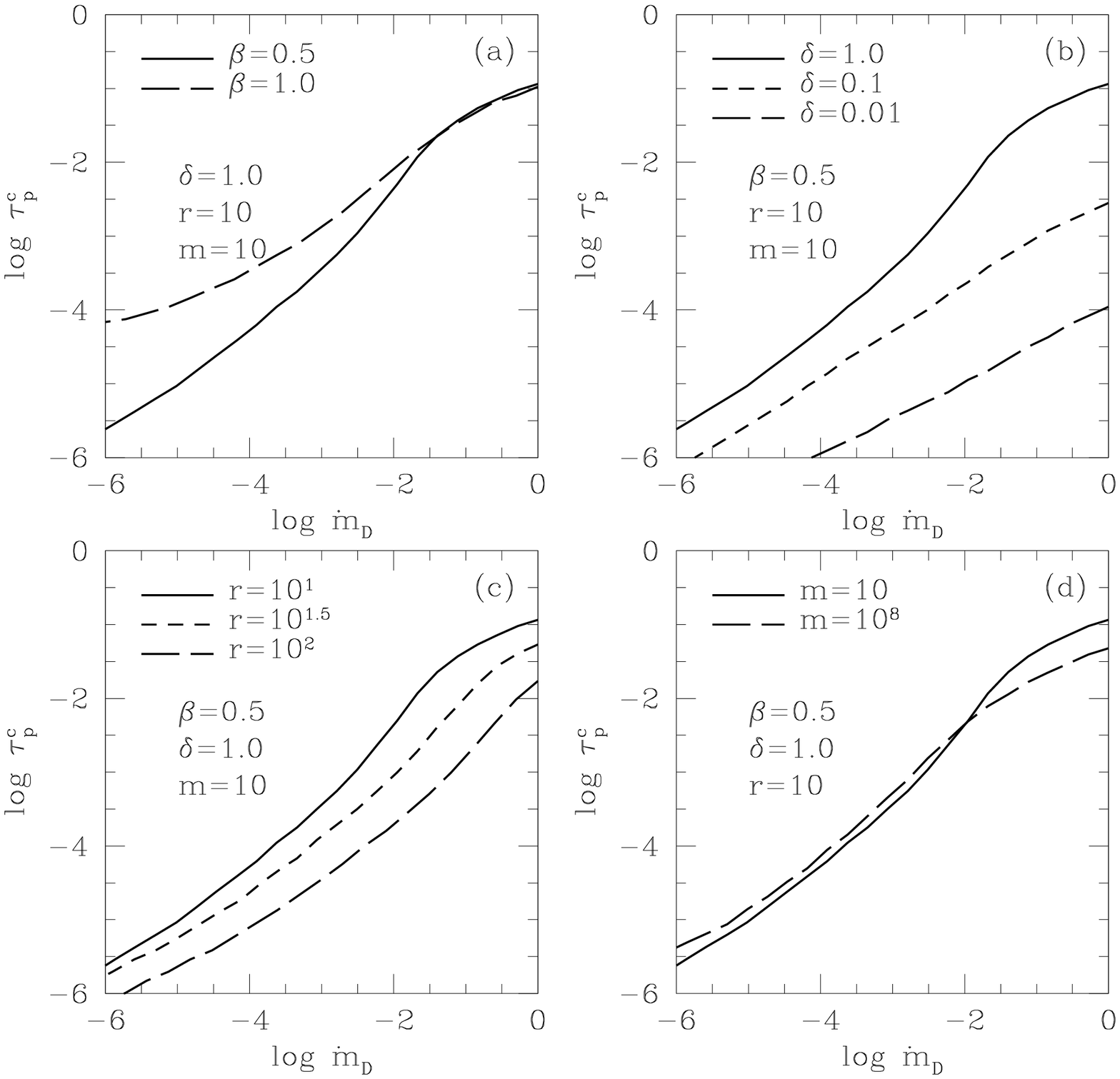} 
\vskip 6.0in 
\caption{\label{taufig} (a) Critical proton optical depth, $\tau_p^c$, plotted as
a function of $\mdot_D$ for a one-temperature corona.  The solid line
was computed for a magnetized corona with $\beta = 0.5$, assuming
$\delta=1$, $r=10$, and $m=10$. The dashed line shows the results for
a non-magnetized corona with otherwise the same parameters.  A
single-temperature corona with $\tau_p < \tau_p^c$ is pair-dominated
($z>1$), while coronae with higher proton optical depth have $z<1$.
(b) Curves of $\tau_p^c$ as a function of $\mdot_D$, calculated for
$m=10$, $r=10$, and $\beta=0.5$, plotted for different values of
$\delta$. 
(c) Curves of $\tau_p^c$ as a function of $\mdot_D$, calculated for
$m=10$, $\delta = 1$, $\beta=0.5$, and values of $r$ as indicated on
the figure. (d) Curves of $\tau_p^c$ as a function of $\mdot_D$,
calculated for different masses of a central black hole.}
\end{figure} 

\begin{figure} 
\includegraphics{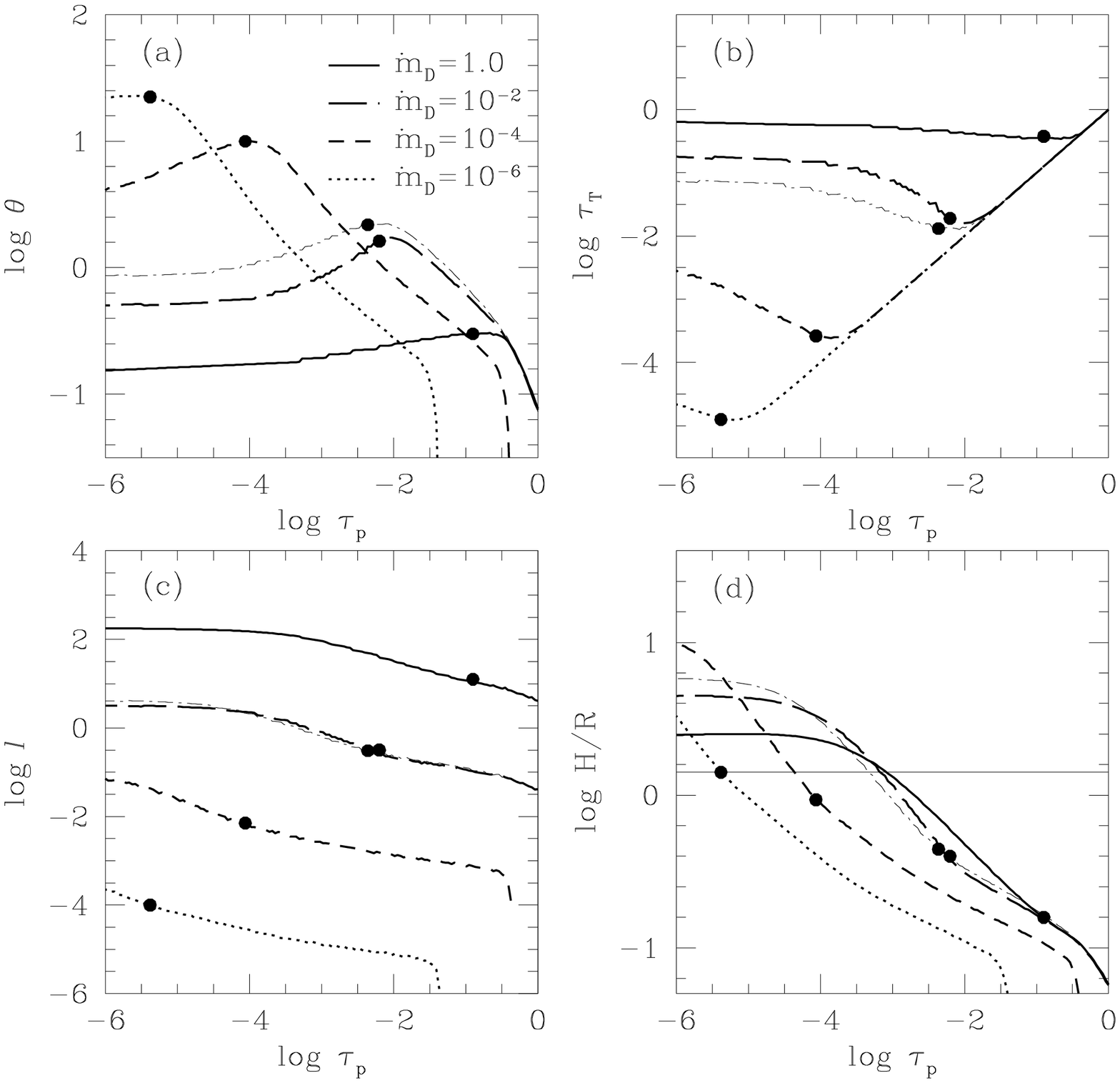}
\vskip 5.7in 
\caption{\label{hfig} On all four panels different curves correspond to
different values of the disk mass accretion rate, $\mdot_D$.  The
other parameters are fixed at $\delta=1$, $r=10$, $m=10$, except for
the dot-dashed curves, which are calculated with $\mdot_D = 10^{-2}$
and $m=10^8$.  The solution with $\tau_p = \tau_p^c$ is marked with a
solid dot on each curve.  (a) Equilibrium temperature, $\theta$, in a
magnetized one-temperature corona plotted as a function of $\tau_p$.
Pair-dominated solutions correspond to $\tau_p \le \tau_p^c$.  (b)
Optical depth for electron scattering, $\tau_T = \tau_p (1+2 z)$,
corresponding to equilibrium corona solutions plotted as a function of
$\tau_p$ for different values of $\dot m_D$.  So long as the corona is
pair-free ($\tau_p \ge \tau_p^c$), $\tau_T$ is equal to the proton
optical depth.  Below $\tau_p^c$, however, pairs dominate the electron
scattering opacity, and $\tau_T$ converges to a constant value. (c)
Dimentionless cooling rate, $\ell$, of the corona plotted as a
function of $\tau_p$ for different values of $\dot m_D$.  (d) Scale
height of the corona determined from equation (\ref{h}).  The thin
solid line corresponds to $H/R=2$; only solutions below this line are
dynamically stable.}
\end{figure}

\begin{figure}
\includegraphics{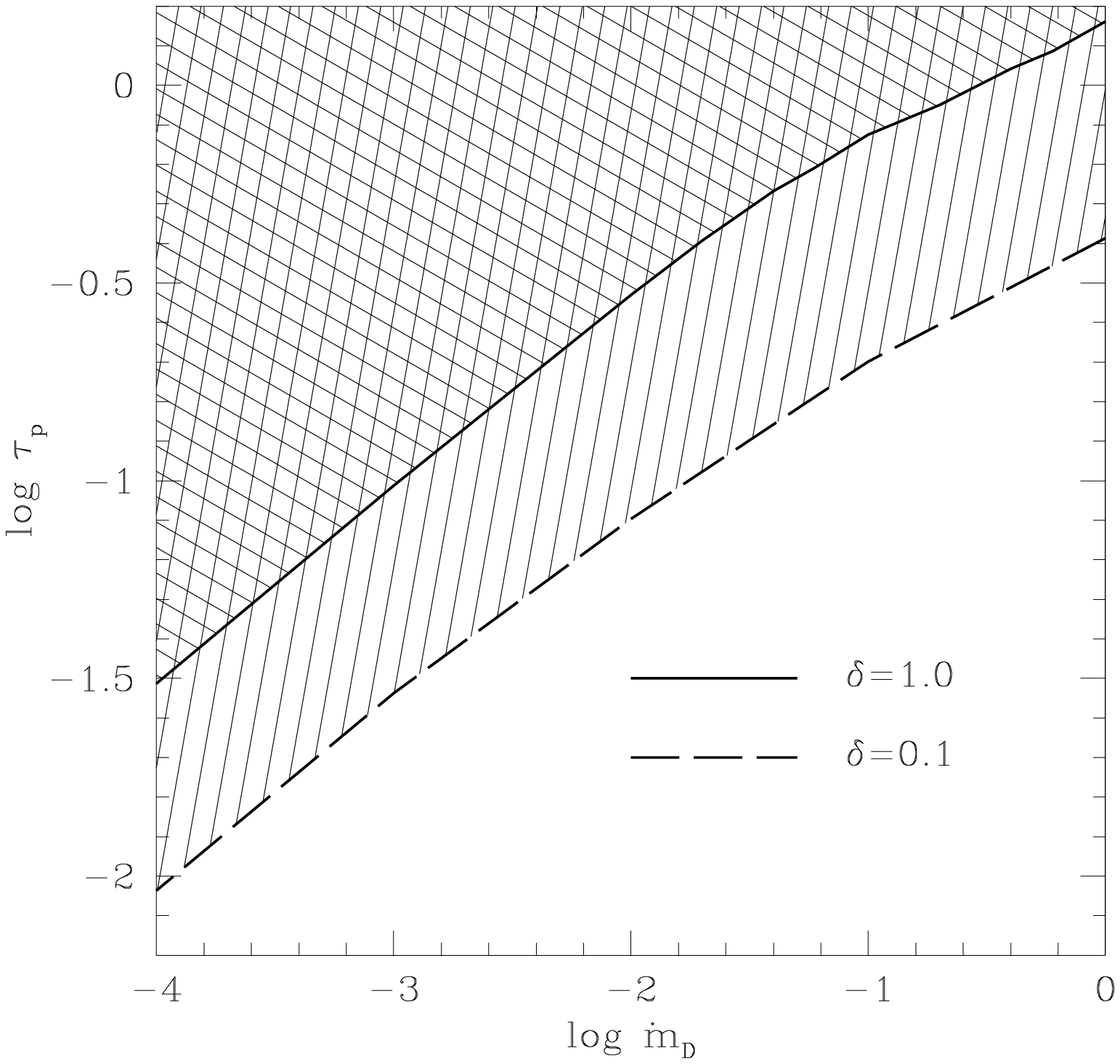} 
\vskip 5.5in 
\caption{\label{twotfig} Regions of the parameter space where
equilibrium solutions for a two-temperature corona are allowed, are
shaded.  The results are shown for $m=10$, $r=10$, $\beta=0.5$ and two
different values of $\delta$, as indicated on the figure.  In the
region below the thick solid line, Coulomb energy transfer rate is too
inefficient to allow the protons in the corona to be in thermal
balance.  A comparison between this panel and Fig.~\ref{taufig}(b)
clearly shows that the shaded regions lie above the $\tau_p^c$ curves
for a one-temperature corona.  This indicates that a two-temperature
corona in thermal equilibrium always has $z\ll 1$.}
\end{figure}

\begin{figure}
\includegraphics{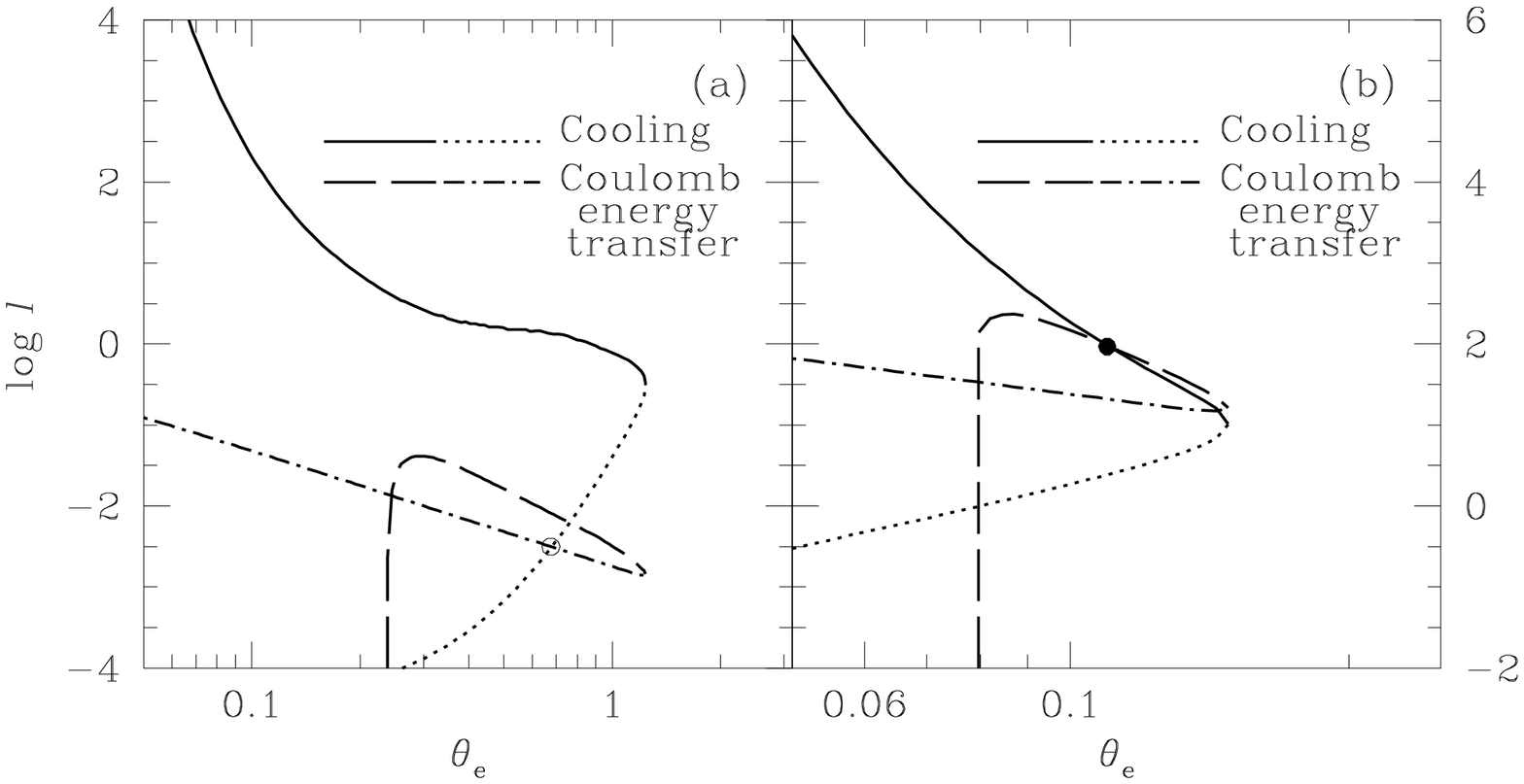} 
\vskip 5.5in 
\caption{\label{adaf1fig} Illustrates the electron energy balance in a
two-temperature accretion flow with a fixed $f$.  Dimensionless
electron cooling rate (compactness) and Coulomb energy transfer rate
in pair equilibrium are plotted vs. electron temperature, $\theta_e$.
Solid and dashed curves show the high-$z$ solution branches for cooling
and heating rates respectively, while dotted and dot-dashed lines show
the low-$z$ branches.  Equation (\ref{ttcore}) is satisfied at the
intersection point. The flow parameters are fixed at $\alpha = 1.0$,
$r=10$, $m=10$, $\beta=0.5$; (a) $f=0.1$, $\mdot= 0.01$; (b) $f=0.03$,
$\mdot=0.14$.  The low-$\mdot$ solution in (a), marked with an open
circle, is pair-free, with $z=2.5\times10^{-5}$; and the high $\mdot$
solution in (b), marked with a solid dot, has a relatively large pair
fraction of $z = 1.9$.}
\end{figure} 

\begin{figure}
\includegraphics{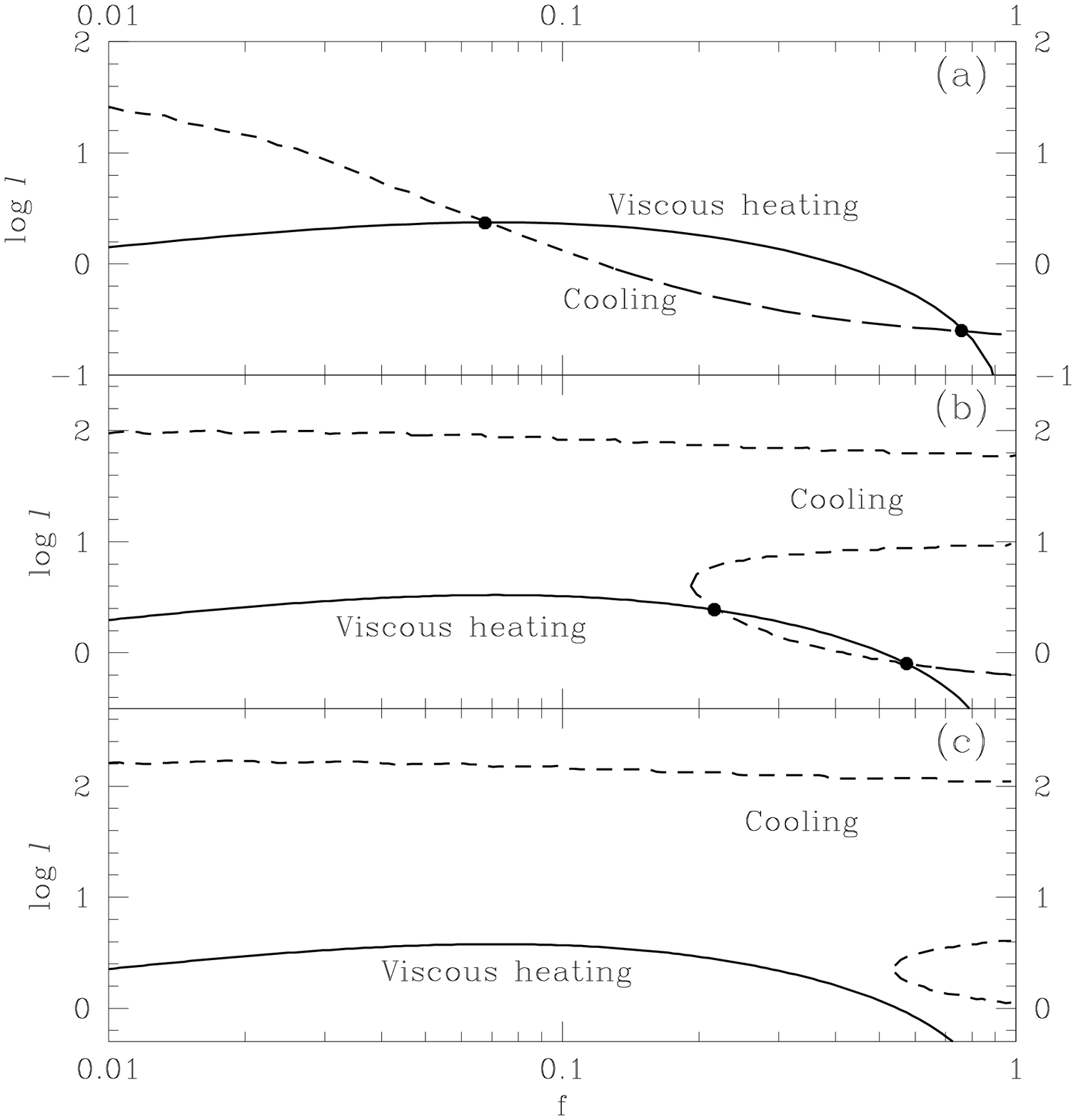} 
\vskip 6.0in 
\caption{\label{adaflfig} Proton energy balance in a two-temperature
accretion flow for $m=10$, $r=10$, $\alpha=1.0$, $\beta=0.5$, and (a)
$\mdot = 0.1$, (b) $\mdot=0.14$, (c) $\mdot=0.16$.  Dimensionless
heating (solid line) and cooling (dashed line) rates for the protons
are plotted as functions of $f$.  Viscous heating is given by the
self-similar accretion flow solutions (Narayan \& Yi 1994).  Cooling
through Coulomb collisions with electrons is obtained by solving the
electron energy balance equation (see Fig.~\ref{adaf1fig}), and is equal to the
local compactness of the gas.  Parts of the cooling curve which
correspond to the high-$z$ branch on Fig.~\ref{adaf1fig} are plotted as a
short-dashed line; the long-dashed line shows the low-$z$ compactness
values.  Solid dots mark the positions of thermal equilibria.  Note
that there are two solutions for $\mdot = 0.1$ and $0.14$ and no
solutions for $\mdot = 0.16$.}
\end{figure} 

\begin{figure}
\includegraphics{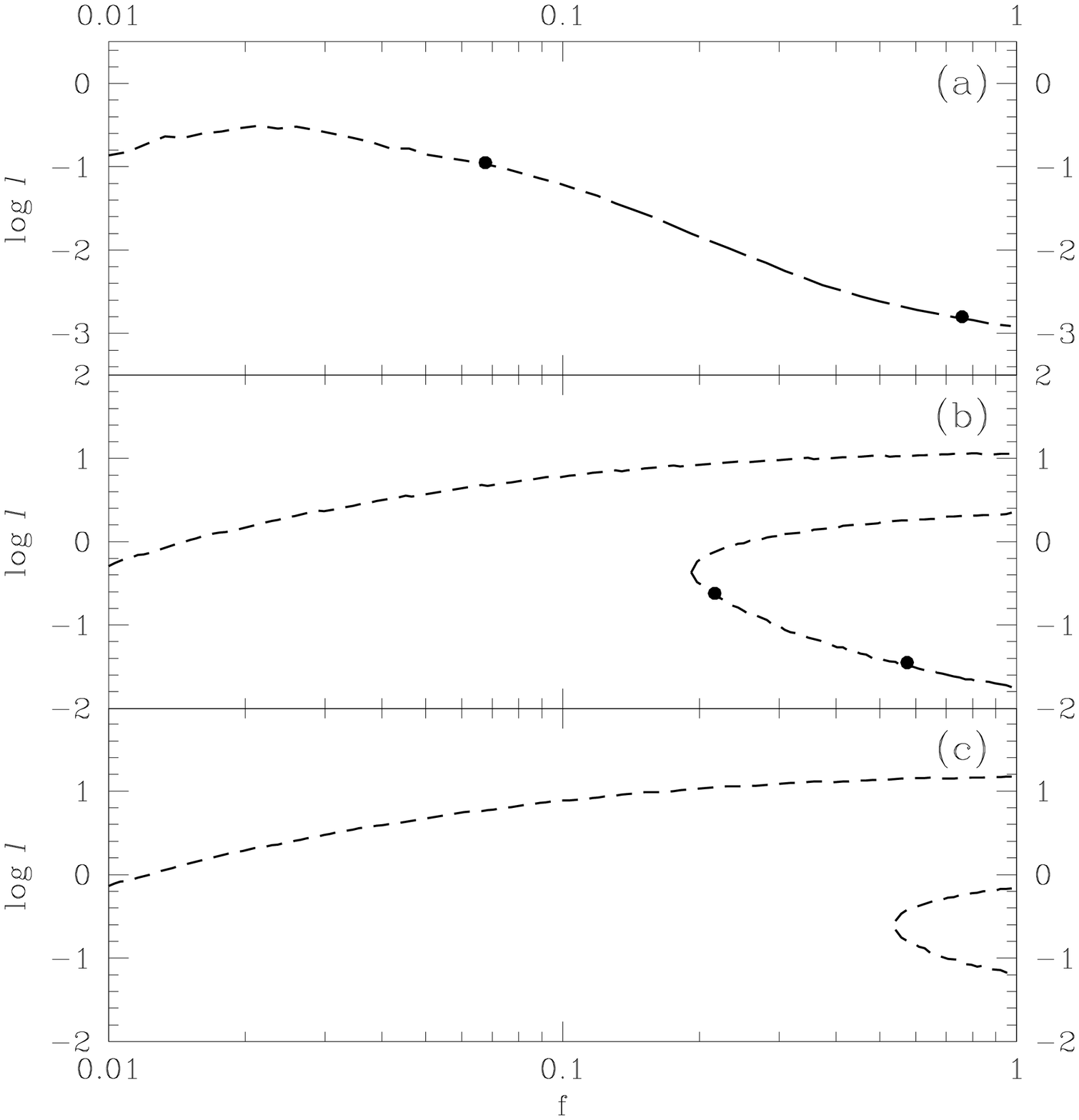} 
\vskip 6.5in 
\caption{\label{adafzfig} Pair fraction $z$ corresponding to the
cooling curves on Fig.~\ref{adaflfig} plotted as a function of $f$ for
the same values of the mass accretion rate: (a) $\mdot = 0.1$, (b)
$\mdot=0.14$, (c) $\mdot=0.16$.  Although there are cooling branches
on panels (b) and (c) that have relatively high pair densities,
thermal equilibria (marked by solid dots), when they exist at all,
invariably have $z <1$.}
\end{figure} 

\begin{figure}
\includegraphics{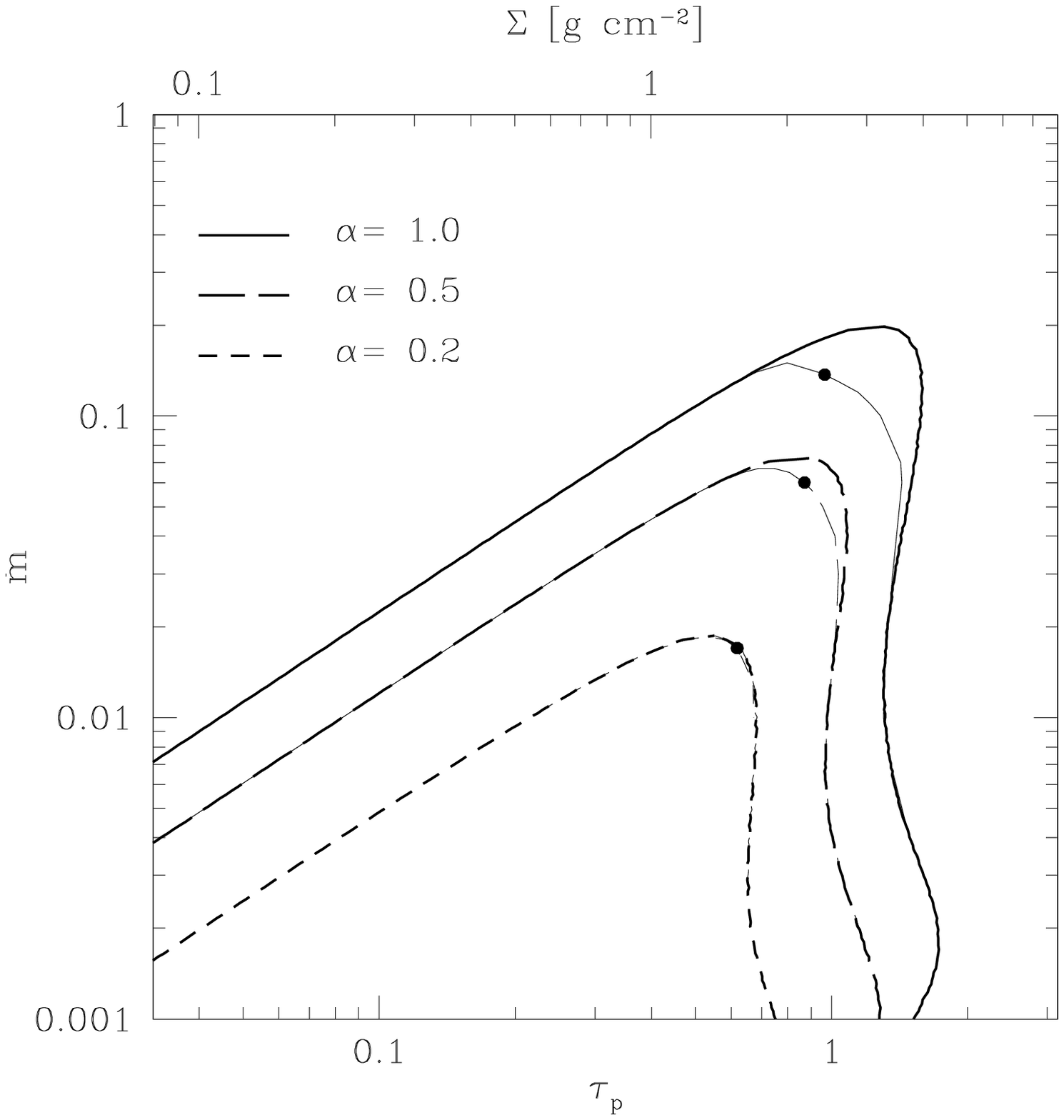} 
\vskip 6.5in 
\caption{\label{sigmafig} Thermal equilibria of optically thin
accretion flows calculated for $m=10$, $r=10$, and $\beta=0.5$..  The
thick curves show the solutions computed without pairs; the thin
curves correspond to the solutions with full account of pairs.  The
maximum pair fraction (indicated by solid dots on the figure) is
$z=0.2$ for $\alpha=1.0$, $z=0.06$ for $\alpha=0.5$, and $z=0.006$ for
$\alpha=0.2$.}
\end{figure} 

\end{document}